\documentclass[12pt]{article}
\usepackage{amsmath,amsfonts, amssymb}
\usepackage{hyperref}
\usepackage{graphicx}
\usepackage{psfrag}
\usepackage[dvips]{color}

\newcommand\sign{\operatorname{sign}}

\newcommand{\address}[1]{\vbox{\center\em#1}}

\newcommand{\ra}{\rightarrow}

\newcommand{\be}{\begin{equation}}
\newcommand{\ee}{\end{equation}}
\newcommand{\ba}{\begin{eqnarray}}
\newcommand{\ea}{\end{eqnarray}}
\newcommand{\bi}{\begin{itemize}}
\newcommand{\ei}{\end{itemize}}

\newcommand{\Tr}{{\rm Tr}}

\newcommand{\Z}{{\mathbb Z}}
\newcommand{\R}{{\mathbb R}}

\newcommand{\HH}{H}
\newcommand{\p}{\partial}

\newcommand{\Ncal}{{\mathcal N}}
\newcommand{\Ocal}{{\mathcal O}}

\newcommand{\Wcal}{{\mathcal W}}

\newcommand{\f}{\frac}
\newcommand{\half}{\frac{1}{2}}
\newcommand{\oo}{\frac{1}}

\newcommand{\aslash}[1]{\,\,{\raise.15ex\hbox{/}\mkern-12mu #1}}
\newcommand{\bslash}[1]{\,\,{\raise.15ex\hbox{/}\mkern-11mu #1}}
\newcommand{\cslash}[1]{\,\,{\raise.15ex\hbox{/}\mkern-10mu #1}}
\newcommand{\dslash}[1]{\,\,{\raise.15ex\hbox{/}\mkern-9mu #1}}

\renewcommand{\bar}{\overline}
\renewcommand{\tilde}{\widetilde}

\renewcommand{\Re}{{\rm Re\,}}

\begin{document}

\bibliographystyle{utphys}

\begin{titlepage}
\rightline{\small{\tt HU-EP-09/29}}
\rightline{\small{\tt NSF-KITP-09-111}}

\begin{center}
\vskip 9mm
{\LARGE
Loop operators and S-duality
\\\medskip\medskip
from curves on Riemann surfaces
}
\vskip 10mm
Nadav Drukker${}^{a}$\footnote{\href{mailto:drukker@physik.hu-berlin.de}
{\rm drukker@physik.hu-berlin.de}},
David R. Morrison${}^{b}$\footnote{\href{mailto:drm@physics.ucsb.edu}{\rm
drm@physics.ucsb.edu}} ,
and
Takuya Okuda${}^{c}$\footnote{\href{mailto:takuya@perimeterinstitute.ca}
{\rm takuya@perimeterinstitute.ca}}
\vskip 5mm
\address{
${}^a$
Institut f\"ur Physik, Humboldt-Universit\"at zu Berlin,\\
Newtonstra\ss e 15, D-12489 Berlin, Germany
}
\address{
${}^b$
Departments of Mathematics and Physics, University of California,\\
Santa Barbara, CA 93106, USA
}
\address{
${}^c$
Perimeter Institute for Theoretical Physics,\\
Waterloo, Ontario, N2L 2Y5, Canada
}
\vskip 6mm

\end{center}

\abstract{
\noindent \normalsize{
We study Wilson-'t~Hooft loop operators in a class of
$\Ncal=2$ superconformal field theories recently
introduced by Gaiotto.
In the case that the gauge group is a product of $SU(2)$ groups,
we classify all possible loop operators in terms of 
their electric and magnetic charges subject to
the Dirac quantization condition.
We then show that this precisely matches Dehn's classification
of homotopy classes of non-self-intersecting 
curves on an associated Riemann surface---the
same surface which characterizes the gauge theory.
Our analysis provides an explicit prediction for the action 
of S-duality on loop operators in these theories which 
we check against the known duality transformation in 
several examples.
 } }

\vfill

\end{titlepage}

\setcounter{footnote}{0}

{\addtolength{\parskip}{-1ex}\tableofcontents}

\section{Introduction}

A new family of interacting four dimensional conformal field 
theories was recently presented in \cite{Gaiotto:2009we}. These 
theories can be motivated from several different points of view:
by using building blocks taken from certain limits of quiver 
theories, in terms of brane webs \cite{Benini:2009gi}, 
or in terms of a dimensional 
reduction of the six dimensional conformal field theory 
with $(2,0)$ supersymmetry describing coincident M5-branes 
wrapped on a Riemann surface.

This Riemann surface plays an important r\^ole in all the different 
descriptions of the theory, as well as its $AdS$ dual 
\cite{Gaiotto:2009gz}. A four 
dimensional conformal field theory exists for any Riemann surface 
(allowing also for certain singularities). There is a one-to-one 
correspondence between the complex structures of the 
surface and the coupling constants of the gauge theory. More 
precisely, a closed surface of genus $g$ corresponds to a 
theory with gauge group $SU(N)^{3g-3}$. Each $SU(N)$ 
factor can have its own coupling and theta angle which 
match the $3g-3$ complex moduli of the surface.%
\footnote{For more details on the case of punctured surfaces
we refer the reader to the original papers 
\cite{Gaiotto:2009we,Gaiotto:2009gz}.}

In addition to the gauge fields (and their superpartners) 
these theories may include fundamental hypermultiplets 
as well as some mysterious strongly interacting conformal 
field theories which have been christened $T_N$. These $T_N$ theories 
have a global $SU(N)^3$ flavor symmetry which will be 
generally gauged by some of the $SU(N)$ factors, coupling them 
to each other. 
The case where all gauge group factors are $SU(2)$ is rather special, 
as the $T_2$ theory is free.

Given that we can continuously deform these theories from weak 
to strong coupling, we would like to understand their behavior 
under S-duality. A particularly useful set of probes for analyzing this 
question are Wilson and 't~Hooft loop operators and
the dyonic mixture of them. They should be mapped to each other 
under the action of S-duality. The purpose of this paper is to classify these 
operators 
and write down their transformation rules under S-duality;  
we find a particularly simple answer for theories 
with a gauge group which is a product of $SU(2)$ factors.

Very recently the question of the behavior of the 
partition function of these conformal field theories under 
S-duality was addressed in \cite{Alday:2009aq}. 
Interestingly, exactly in this case based on $SU(2)$ groups these authors
found that the partition function of the four dimensional theory 
is equal to a correlation function of Liouville theory on the 
aforementioned Riemann surface.%
\footnote{A proposal for theories with $N>2$ was given in 
\cite{Alday:2009aq,Wyllard:2009hg}.} 
We expect a generalization 
of their prescription to apply also to the calculation of the expectation 
values of loop operators. As we became aware of the great 
interest in understanding loop operators in these theories, 
we decided to publish our $SU(2)$ classification now, and defer to 
future work \cite{N>2} some
obvious open questions such as the analogue of these operators
in Liouville theory, or 
the general case based on $SU(N)$ groups.

A loop operator may have an arbitrary shape in space-time, but in 
order to classify the possible {\it types} of operators it makes sense to 
choose a particular geometry for the curve. The simplest choices 
are a straight line or a circle (which are related by a conformal 
transformation). Locally, any smooth loop is approximately straight, 
and in these particular cases the loop operator can also preserve 
global supersymmetry. 

As may be expected, after fixing the geometry of the loop in 
space-time, the remaining degrees of freedom are related to its 
gauge structure. Since this is intimately related to the associated Riemann 
surface \cite{Gaiotto:2009we}, we expect to be able to classify loop
operators in terms of the geometry of the surface. In fact, the classification
turns out to be simple and beautiful:
the loop operators are in one-to-one correspondence with 
non-self-intersecting curves on the surface (up to homotopy).%
\footnote{%
The gauge theory endows the surface only 
with a complex structure. One can always find a surface with a 
hyperbolic metric and the same complex structure. Using this 
metric is sometimes convenient, as each homotopy class has 
a unique geodesic representative (and if the homotopy class can 
be represented by a non-self-intersecting curve,  the 
geodesic is also non-self-intersecting). Therefore this matching applies also 
to the classification of non-intersecting 
geodesics with respect to the hyperbolic metric.}
To avoid confusion we repeat that the loops have 
a fixed geometry in space-time. The curves live on an 
auxiliary surface which is useful in order to classify these field 
theories.

This correspondence can then be used to understand the action of 
S-duality on the loop operators. It is believed that the S-duality 
group is isomorphic to the mapping class group
of the surface \cite{Witten:1997sc,Halmagyi:2004ju,Gaiotto:2009we}.
The action of this group on the non-self-intersecting
curves is quite complicated, 
but is well understood. This provides therefore a prediction for 
the action of the S-duality group on arbitrary Wilson-'t~Hooft loop 
operators in the generalized quiver theories with $SU(2)$ gauge 
groups.

As motivation for the relation between curves and loop operators, 
it is useful to keep in mind the realizations of these field theories in 
M-theory. First, these theories arise on coincident M5-branes 
wrapping a Riemann surface. The loop operators correspond to 
M2-branes ending on the M5-branes along a 2-surface with one 
direction on the Riemann surface (and the other in the remaining 
four flat directions). For a large number of M5-branes 
there is a dual description of this system within the $AdS$/CFT 
correspondence \cite{Maldacena:2000mw,Gaiotto:2009gz}. 
In the dual geometry this Riemann surface also plays a r\^ole 
and the loop operators are again described by M2-branes 
with two directions inside $AdS_5$ and the third in the compact 
space. In the supergravity dual the supersymmetric 
embeddings are given by arbitrary geodesics 
on the Riemann surface with hyperbolic metric, 
allowing also for self-intersections (see Appendix~\ref{app-ads}). 
From the M-theory point of view the restriction to non-self-intersecting loops is 
quite mysterious and deserves further exploration.

The paper is organized as follows.
In Section~\ref{sec-NF=4}, we illustrate our analysis by 
focusing on the prototypical conformal gauge theory with $\Ncal=2$
supersymmetry, namely the $SU(2)$ gauge theory with $N_F=4$ flavors.
Following this, Section~\ref{sec-CFT} discusses
Wilson-'t~Hooft loops in an arbitrary generalized quiver theory with $SU(2)$
gauge group factors characterized by a punctured Riemann surface.
In Section~\ref{sec-geodesics} we describe the classification
of homotopy classes of curves 
with no self-intersection on the Riemann surface.
The data used for the gauge theory and topological classifications 
are shown to be identical.  
Furthermore we discuss the transformation rules of curves 
under the action of the mapping class group. 
In cases where the S-duality transformations
are explicitly known, we show that they 
agree with the action of the mapping class group. 
In other cases the geometric analysis provides a prediction for the 
action of S-duality.
We summarize our results and present some 
further discussions in Section~\ref{sec-discussion}.

In Appendix \ref{app-ads} we analyze the supersymmetric embeddings 
of M2-branes in the Maldacena-Nu\~nez geometry. This is the supergravity 
analogue of the calculation in the body of the paper but in a very 
different regime, applicable for theories based on $SU(N)$ 
with large $N$, rather than $SU(2)$. 
This appendix can be read independently of the rest of the paper.

\section{Prelude:  \texorpdfstring{$SU(2)$}{SU(2)} gauge theory with 
 \texorpdfstring{$N_F=4$}{NF=4}}
\label{sec-NF=4}

As a first example of a conformal $\Ncal=2$ theory with $SU(2)$ factors
we consider the simple case of a single $SU(2)$ gauge group and
$N_F=4$ hypermultiplets in the fundamental representation. 
 We want 
to analyze the supersymmetric loop operators of this theory. As mentioned 
in the introduction, the geometry of the loop will always be a straight line 
or a circle and the classification of operators corresponds to the 
charges they carry.

The most well-known loop operators are Wilson loops 
\cite{Wilson:1974sk}. The supersymmetric generalization in $\Ncal=4$ 
super Yang-Mills theory was introduced in \cite{Rey:1998ik,Maldacena:1998im}.
Physically, these operators correspond to the insertion of an electrically 
charged BPS particle with infinite mass along the loop. 
The construction involves coupling the Wilson loop also to a
real scalar field, which for supersymmetry should be in the same
multiplet as the gauge field. 
In theories with extended $\Ncal=2$ supersymmetry the 
vector multiplet includes a complex scalar $\phi$ and we can take
the real field $\Re[\phi]$ (see \cite{Zarembo:2002an}).

Wilson loop operators are defined as the trace of the holonomy along a loop%
\footnote{All fields have canonical kinetic terms and hence the explicit gauge 
coupling in the definition.}
\ba
\Tr_R\, {\cal P} \exp\left[ g_\text{YM}\oint (i  A+\Re[\phi]\,ds)\right],
\label{WL}
\ea
where the parameter $s$ is normalized so that $|dx^\mu/ds|^2=1$.
For a given geometry (for us a line or a circle) they are labeled by 
representations $R$ of the gauge group $G$. In the case of a single $SU(2)$ 
this is a single positive half-integer spin $j=q/2$.

't~Hooft loops are the magnetic counterpart of
Wilson loops, and they insert probe monopoles along a 
loop in space-time.
In the original definition by 't~Hooft \cite{'tHooft:1977hy},
the magnetic loop operators are classified by their topological charge.
Kapustin introduced a finer classification
\cite{Kapustin:2005py} allowing for topologically trivial loops:
a supersymmetric 't~Hooft loop is defined by performing the 
path integral over field configurations with a specific singularity 
along the curve. Using spherical coordinates $(r,\vartheta,\varphi)$ in 
the space transverse to 
the time-like line the singularity near the loop takes the form%
\footnote{It is necessary to excite the scalar field
in order to preserve some supersymmetries.
The  particular phase $-i$ is needed
for the 't~Hooft loop to preserve the same supercharges as
the Wilson loop (\ref{WL}). This expression is modified 
for non-zero theta-angle.}
\ba
g_\text{YM}A=\f{\mu}2(1-\cos\vartheta)d\varphi
+\Ocal(1/r)\,,
\qquad
g_\text{YM}\phi=-i\f{\mu}{2r}+\Ocal(1)\,,
\label{t-background}
\ea
with
\ba
\mu=\left(
\begin{array}{cc}
p/2&0\\
0&-p/2
\end{array}
\right).
\label{t-background-Nf4}
\ea
The loop operators with odd $p$ (which would be topologically non-trivial for 
an $SO(3)$ gauge group)
can be defined only in a theory where none of the fields 
are charged under the center $\Z_2$  of the gauge group. 
Such is the case for a theory with matter only in the adjoint 
representation, but the case we consider here has four hypermultiplets 
in the fundamental representation, for which the usual Dirac 
quantization condition applies, meaning that $p$ should be 
an even integer.

Supersymmetric dyonic loops are defined
by inserting Wilson loops for the gauge group
that remains unbroken in the 't~Hooft loop background
(\ref{t-background}). For the Wilson loop carrying $q$ 
units of electric flux we should rescale the
singularity for the scalar field $\phi$ in (\ref{t-background})
by $\sqrt{1+q^2/p^2}/2$.%
\footnote{See \cite{Kapustin:2005py} 
and \cite{Kapustin:2006pk} for
more details.}

Thus a general loop operator is specified by two integers: $p$, which 
is even, is the magnetic charge and $q$ is the electric charge. 
Two pairs of charges which are related by the common Weyl group action
\ba
(p,q)\sim (-p,-q)
\ea
give identical loop operators.
We may therefore assume $p\geq 0$ and for $p=0$ we can take 
$q\geq 0$.
As the simplest example, the Wilson loop in the fundamental representation
has weights $(0,1)$.

\begin{figure}[ht]
\begin{center}
\begin{tabular}{cccc}
\includegraphics[scale=.5]{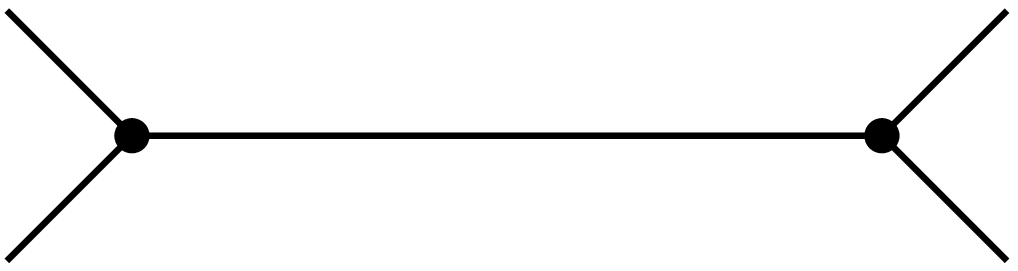}
&&
\psfrag{gamma}{$\gamma$}
\psfrag{delta}{$\delta$}
\includegraphics[scale=.5]{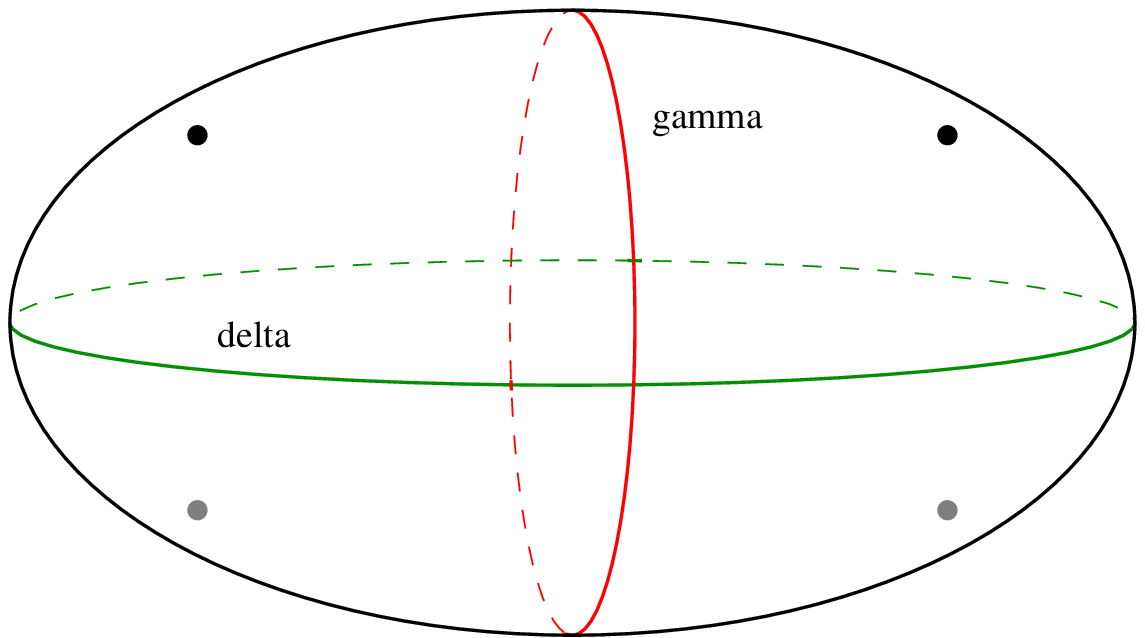}
\\
(a)&&(b)
\end{tabular}
\parbox{5in}{
\caption{
(a) A ``quiver'' diagram that represents the $SU(2)$ gauge theory
with $N_F=4$ flavors.
(b)
A sphere with four punctures obtained by fattening the
quiver diagram. The geodesics we discuss are
for the hyperbolic metric where each puncture has a $2\pi$ deficit 
angle and is
at the end of an infinite tube.
The red curve represents $\gamma$ represents the Wilson loop
in the fundamental representation, while the green curve $\delta$
corresponds to the minimal 't~Hooft loop.
\label{sphere-4pts}}}
\end{center}
\end{figure}

In anticipation of a later generalization, it is convenient to represent 
the gauge theory by a version of a quiver diagram
shown in Figure~\ref{sphere-4pts}(a), where the 
internal edge is the $SU(2)$ gauge group and 
each open edge is a flavor group. The vertices are hypermultiplets in 
the fundamental of the both $SU(2)$ flavor groups as well as 
the gauge group.

It was observed in \cite{Witten:1997sc} that the parameter space 
of gauge couplings in this theory coincides with the Teichm\"uller space
of a four-punctured sphere obtained by fattening the quiver diagram,
as shown in Figure~\ref{sphere-4pts}(b).%
\footnote{More precisely, the observation was that
the parameter space modulo
S-duality coincides with the complex structure
moduli space of a sphere with four equivalent punctures.
If we identify the mapping class group with the
S-duality group, the two statements are equivalent.}

From this point of view, the quiver diagram Figure~\ref{sphere-4pts}(a) 
arises from a specific choice of ``pants decomposition'' of the four-punctured 
sphere which is natural at a corner of moduli space. 
The closed curve $\gamma$ separates two punctures from the other 
two and each half of the sphere is topologically a ``pair of pants'' (or 
a three-punctured sphere). The curve $\delta$ gives another choice 
of quiver diagram describing the same gauge theory in a different 
S-duality frame.

In terms of this geometry we identify all supersymmetric loop operators with 
homotopy classes of non-self-intersecting curves on a four-punctured sphere 
as follows:
closed curves without self-intersections are classified (homotopically) 
by the number of times they cross the curve $\gamma$ and the twist 
they perform along $\gamma$. That is a pair of integers $(p,q)$ 
where the number of crossings $p$ is even and positive and 
$q$ arbitrary (if $p=0$, then the sign of $q$ is ill defined and it 
will be taken to be positive).

For example, $\gamma$ itself corresponds to $(0,1)$, and the curve
consisting of $q$ 
copies of $\gamma$ is represented by $(0,q)$. The curve $\delta$ 
is labeled by $(2,0)$. By cutting the surface along $\gamma$ twisting 
it and re-gluing, we get all curves with labels $(2,q)$.
According to Dehn's theorem explained in
Section~\ref{sec-geodesics}, all homotopy classes with non-self-intersecting 
curves are classified by this labeling.

We see that  Wilson-'t~Hooft operators
and non-self-intersecting curves
are in one-to-one correspondence since they are both
labeled by $(p,q)$ with the same identification.
As we will see in Sections~\ref{subsec-gauge-examples} and 
\ref{subsec-geodesics-examples}, this classification is consistent with the
identification of the S-duality group $SL(2,\Z)$
with the mapping class group of the four-punctured sphere.

It is also natural to consider open curves connecting the punctures 
on the sphere. We discuss them in the next section after presenting the 
general case.

\section{Classification of loop operators in gauge theory}
\label{sec-CFT}

As discussed earlier, new $\Ncal=2$ superconformal
theories were discovered in \cite{Gaiotto:2009we}.
These are ``generalized quiver theories'' which involve the
conformal theories denoted by $T_N$ in \cite{Gaiotto:2009gz},
as well as $SU(N)$ gauge groups with various values of $N$.
For $N>2$, $T_N$ is an exotic theory without a known
Lagrangian description, while
$T_2$ is simply a free theory that contains four
hypermultiplets, or equivalently eight
half-hypermultiplets.
The theory has
flavor symmetry group $SU(2)\times SU(2)\times SU(2)$
under which 
all fields 
transform in the
$({\bf 2,2,2})$ representation.

\begin{figure}[ht]
\begin{center}
\begin{tabular}{ccccc}
\includegraphics[scale=.4]{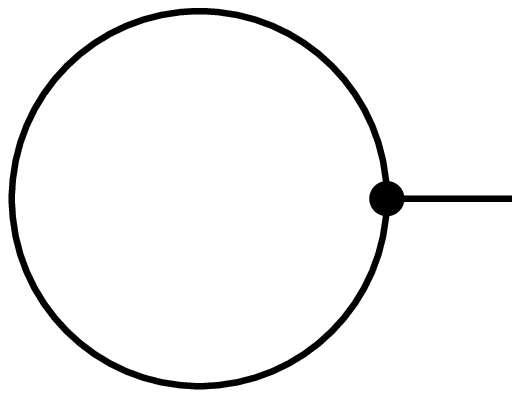}
&\qquad&
\includegraphics[scale=.4]{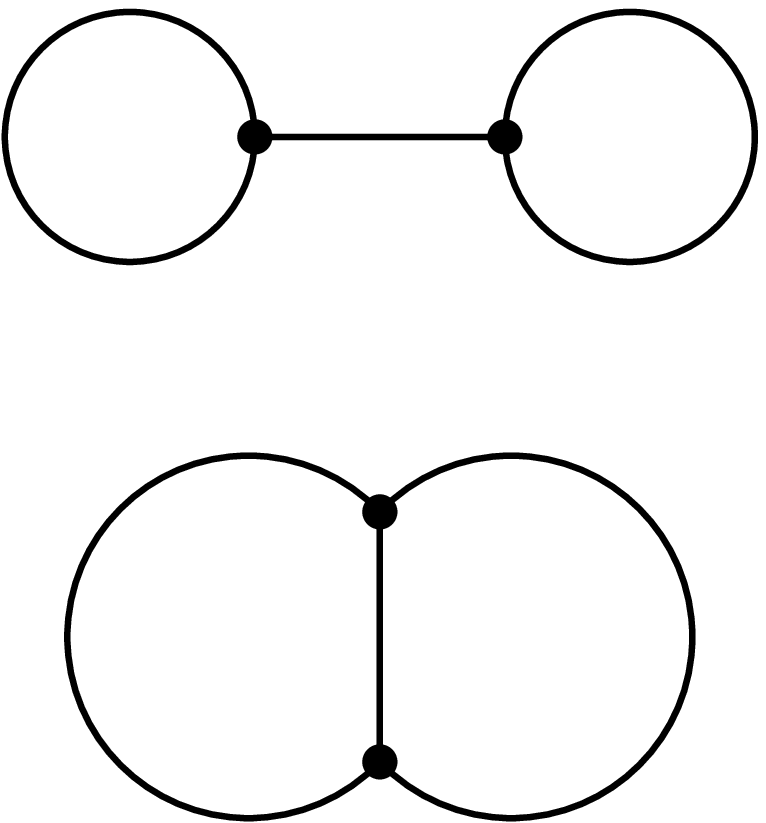}
\\
(a)&&(b)
\end{tabular}
\parbox{5in}{
\caption{Examples of $SU(2)$ quiver diagrams.
(a) The $\Ncal=2^*$ gauge theory (a mass deformation of $\Ncal=4$ SYM) 
corresponds to a once-punctured torus.  
(b) Two quiver gauge theories that are dual to each other.
They  correspond to a genus two surface with no punctures.
\label{fig:SU(2)quivers}}}
\end{center}
\end{figure}

In this paper we focus on a subclass of generalized
quiver theories
which are based on $SU(2)$ gauge groups
and $T_2$.  Such a theory is represented by a generalized quiver
diagram built  from  trivalent vertices
connected by edges.  See Figure~\ref{sphere-4pts}(a)
as well as Figure~\ref{fig:SU(2)quivers} for examples.
An internal edge represents an $SU(2)$ factor
in the gauge group whereas each $T_2$ theory 
corresponds to a trivalent vertex. The external (open) edges 
correspond to an $SU(2)$ flavor symmetry.

By thickening the edges, we can associate to each quiver
diagram the topology of a Riemann surface.
An open leg of a trivalent vertex corresponds to a puncture on the surface.
Let $g$ denote the genus of the surface and $n$ the number
of punctures.%
\footnote{The number of trivalent vertices is $2g-2+n$, the number of 
internal edges is $3g-3+n$ and the number of open edges is $n$.}
Then the theory has gauge group
\ba
G=SU(2)^{3g-3+n}=\prod_{j=1}^{3g-3+n}SU(2)_j
\ea
and flavor symmetry group
\ba
SU(2)^n=\prod_{j=3g-2+n}^{3g-3+2n} SU(2)_j.
\ea
\begin{figure}[t]
\begin{center}
\begin{tabular}{cccc}
\includegraphics[scale=.7]{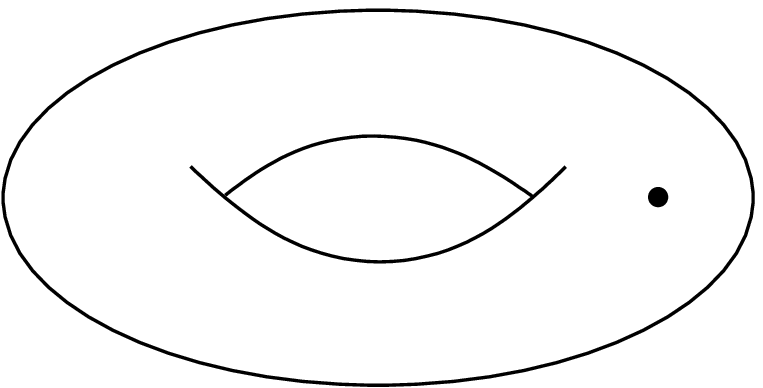}
&\qquad&
\includegraphics[scale=.7]{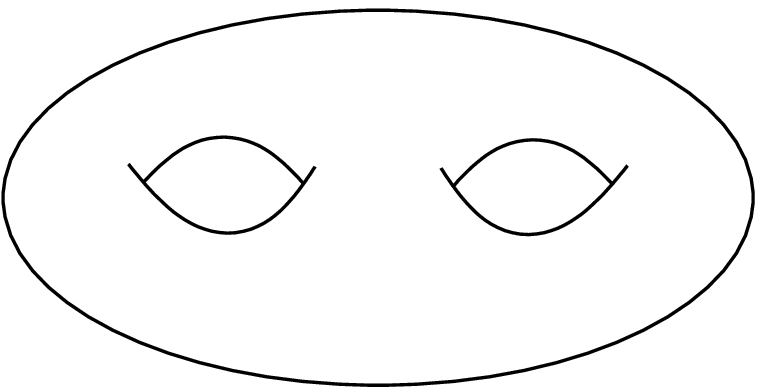}
\\
(a)&\qquad&(b)
\end{tabular}
\parbox{5in}{\caption{(a) A once-punctured torus.
(b) A genus two Riemann surface.
Two S-dual realizations of the theory based on 
this surface are given in Figure~\ref{fig:SU(2)quivers}(b).
\label{surfaces}}}
\end{center}
\end{figure}

Our aim is to classify Wilson-'t~Hooft loop operators in this theory. 
In Section~\ref{sec-NF=4} we presented these objects for a theory 
whose gauge group is a single $SU(2)$. For a general gauge group 
the Wilson loops are again given by (\ref{WL}), where now $R$ is an 
arbitrary representation of the gauge group. 
The representation $R$ may be specified by
their highest weights $\nu\in \Lambda_{\rm w}$, where
$\Lambda_{\rm w}$ is the weight lattice.
Supersymmetric 't~Hooft loops are given again by (\ref{t-background}) 
where $\mu$ generally takes values in the coweight lattice $\Lambda_{\rm cw}$
defined as the dual of the root lattice.

The most general supersymmetric loop operator will be dyonic,
and this requires first choosing a magnetic source and adding on top an arbitrary 
Wilson loop in the gauge group left unbroken by the magnetic 
background.\footnote{See \cite{Kapustin:2005py} 
and \cite{Kapustin:2006pk} for
a more precise definition of a dyonic loop operator.}
It was argued in \cite{Kapustin:2005py}
that general supersymmetric Wilson-'t~Hooft operators
are labeled by a pair of magnetic and electric weights
up to identification by the action of the Weyl group:
\ba
(\mu,\nu) \in \Lambda_{\rm cw}\times \Lambda_{\rm w},~~~~
(\mu,\nu)\sim(w \cdot \mu, w\cdot \nu),~~~w\in \Wcal.
\ea

As mentioned for the example in Section~\ref{sec-NF=4}, in the 
presence of matter fields charged under the center of the gauge 
group, some restrictions need to be imposed 
on the allowed magnetic weights. 
The singular background (\ref{t-background})
has a Dirac string along $\vartheta=\pi$ (where $\vartheta$ is an
angular variable in the transverse space).
When one goes around the Dirac string, a matter field $\Phi$
transforms as
\ba
\Phi\ra e^{2\pi i \mu}\cdot \Phi,
\ea
where the exponential acts on $\Phi$ according to the representation
 in which $\Phi$ transforms.
The Dirac string is thus invisible if and only if
$\exp(2\pi i\mu)$ acts trivially on all matter fields.
The set of allowed magnetic weights $\mu$ that satisfy
this Dirac condition forms a sublattice of $\Lambda_{\rm cw}$.

This analysis of Wilson-'t~Hooft operators applies to any $\Ncal=2$ 
gauge theory. We now wish to apply it to the specific theories we 
are considering, where the gauge group is a product of $SU(2)$ 
factors. The general weights parameterizing the loop operator 
are the set of integers
\ba
(p_1,p_2,\ldots,p_{3g-3+n}; q_1,q_2,\ldots,q_{3g-3+n}).
\ea
The Weyl group $\Wcal=(\Z_2)^{3g-3+n}$ 
acts by $p_j \ra -p_j,q_j\ra -q_j$ for any $j$ 
and sets of integers related in this way 
specify the same operator.
The definition of
this operator involves two steps.
First we demand that
the path-integral be over singular field configurations
satisfying near the loop%
\footnote{For simplicity we wrote the expression for the case when the 
complexified gauge coupling $\tau=\theta/2\pi+4\pi i/g_\text{YM}^2$ is purely 
imaginary. A generalization exists for any theta angle.}
\be
\begin{aligned}
g_\text{YM}^{(j)}A^{(j)}&=
\begin{pmatrix}
p_j/2&0\\
0&-p_j/2
\end{pmatrix}
\f{1-\cos\vartheta}2
d\varphi+\Ocal(1)\,,
\\
g_\text{YM}^{(j)}\phi^{(j)}&=
-\f{i}{2r}
\begin{pmatrix}
\omega_j/2&0\\
0&-\omega_j/2
\end{pmatrix}
+\Ocal(1)\,.
\label{t-background-SU(2)}
\end{aligned}
\ee
If $p_j$ is non-zero, $SU(2)_j$ is 
broken to its maximal torus $U(1)$
by the background fields in (\ref{t-background-SU(2)}).
Second the path-integral is performed
with the insertion of a Wilson loop specified
by $q_j$ for the unbroken gauge group.
The coupling $\omega_j$ will be chosen so that the loop operator
preserves the same supercharges as the pure Wilson loop (\ref{WL}).
Though this definition is sufficient for classification purposes,
a more precise one should include regularization
and boundary terms.

Supercharges preserved by a loop operator
depend on its magnetic charge, its electric charge and in addition 
the choice of $\omega_j$. Thus $\omega_j$ should be adjusted 
independently for each $SU(2)_j$ such that the supercharges 
are shared by the full operator.
A heuristic  way to determine the value of $\omega_j$ is 
by considering the classical field configuration 
produced by the dyonic operator with weights $(p_j,q_j)$. 
Generalizing the treatment of the Abelian case in \cite{Kapustin:2005py}, 
the gauge potential and scalar field take the form
\be
\begin{aligned}
g_\text{YM}A_{\rm cl}&=
i\begin{pmatrix}
q_j/2&0\\
0&-q_j/2
\end{pmatrix}
\f{dt}{2r}
+
\begin{pmatrix}
p_j/2&0\\
0&-p_j/2
\end{pmatrix}
\f{1-\cos\vartheta}2
d\varphi,
\\
g_\text{YM}\phi_{\rm cl}&=
-\f{i}{2r}
\begin{pmatrix}
\omega_j/2&0\\
0&-\omega_j/2
\end{pmatrix}.
\end{aligned}
\ee
The gaugino variation vanishes when $|\omega_j|=\sqrt{p_j^2+q_j^2}$. In 
general we get a similar condition for each $SU(2)$ factor and 
they all have to be consistent, so that if for a purely electric loop operator 
we take a real scalar field with $\omega_j=i$ 
then for a pair of magnetic and electric weights 
$(p_j,q_j)$, we find $\omega_j=p_j+i q_j$.  Thus we require that for all $j$
\be
\omega_j=p_j+i q_j\,.
\ee

It turns out that in the particular class of
quiver theories we are considering,
it is rather natural to generalize the notion of Wilson-'t~Hooft loop operators.
This is done by introducing non-dynamical gauge and scalar fields
$\hat A^{(j)}$ and $\hat \phi^{(j)}$ for the flavor symmetry groups $SU(2)_j$
and  setting them to
\ba
\hat A^{(j)}= \left(
\begin{array}{cc}
p_j/2&0\\
0&-p_j/2
\end{array}
\right)\f{1-\cos\vartheta}2
d\varphi,\qquad
\hat\phi^{(j)}=
-\f{i}{2r}
\begin{pmatrix}
p_j/2&0\\
0&-p_j/2
\end{pmatrix},
\label{t-background-flavor}
\ea
for $j=3g-2+n,\dots,3g-3+2n$.

We are using the same notation $p_j$
both for dynamical and non-dynamical fields since this will allow
us to treat them uniformly.
All dynamical fields charged under the flavor groups $SU(2)_j$
couple to these background gauge fields and therefore are sections
of appropriate vector bundles. The non-dynamical scalar is required if 
we want to preserve the supersymmetry of the hypermultiplets it 
couples to.

One reason the excitation of non-dynamical fields is natural is that they 
arise when we consider 
these theories as limits of theories with additional $SU(2)$ gauge 
factors, which reduce to flavor groups in the decoupling limit. If 
there was a non-trivial bundle, this remains in the decoupling 
limit. Due to that, it is also natural to identify 
non-dynamical field configurations related by 
the (flavor) Weyl group action $p_j\to -p_j$. In any case, since now the 
Weyl group is a global symmetry, its action can be read from the 
way the fields are charged under it (and can also be absorbed 
by a field redefinition).

The introduction of such non-dynamical gauge fields 
is a deformation of the theory rather than an operator.
Such a deformation, combined with the Wilson-'t~Hooft operator
above, defines a generalized Wilson-'t~Hooft loop.
Our discussion so far implies that generalized Wilson-'t~Hooft loops
are labeled by $6g-6+3n$ integers
\ba
(p_1,p_2,\ldots,p_{3g-3+2n};
q_1,q_2,\ldots, q_{3g-3+n})
\label{gen-weights}
\ea
subject to identification by the (independent) Weyl group actions
$(p_j,q_j)\ra (-p_j,-q_j)$ for $j=1,2,\ldots,3g-3+n$ as well as 
$p_j\ra -p_j$ for $j=3g-2+n,\dots,3g-3+2n$. 
Ordinary Wilson-'t~Hooft loop operators correspond to 
generalized weights 
such that
$p_{3g-2+n}=\ldots=p_{3g-3+2n}=0.$

Let us now revisit the Dirac condition on 
the set of allowed Wilson-'t~Hooft loops discussed above.
Each trivalent vertex has matter fields
which transform in the $({\bf 2,2,2})$
representation under $SU(2)_i\times SU(2)_j \times SU(2)_k$,
where each of $i, j$ and $k$ labels
either an edge between two vertices
or an open leg attached to the trivalent vertex.
We allow the possibility that two of $i,j$ and $k$ are identical.
When one transports such a matter field $\Phi$ 
around a Dirac string in the background
fields (\ref{t-background-SU(2)}) and (\ref{t-background-flavor}),
it gets transformed by the three matrices in the defining representations
\ba
\Phi \ra 
\left(
\begin{array}{cc}
e^{\pi i p_{i}}&0\\
0&e^{-\pi ip_{i}}
\end{array}
\right)
\otimes
\left(
\begin{array}{cc}
e^{\pi i p_{j}}&0\\
0&e^{-\pi ip_{j}}
\end{array}
\right)
\otimes
\left(
\begin{array}{cc}
e^{\pi i p_{k}}&0\\
0&e^{-\pi ip_{k}}
\end{array}
\right)
\cdot \Phi.
\ea
For $\Phi$ to be single-valued, we thus require that
\ba
p_{i}+p_{j}+p_{k}\in 2\Z.
\label{gen-even}
\ea

This leads to the main result of this section.
Generalized Wilson-'t~Hooft loops
in the $\Ncal=2$ conformal generalized quiver theory,
 corresponding to a Riemann surface of genus $g$ with $n$ punctures,
are labeled by the magnetic and electric weights
\ba
(p_j,q_j)\equiv (p_1,p_2,\ldots,p_{3g-3+2n};
q_1,q_2,\ldots, q_{3g-3+n})
\label{gen-weights2b}
\ea
where due to the action of the Weyl group we may assume that 
$p_j\geq0$ and for any $j$ such that $p_j=0$, we can take $q_j\geq0$. 
In addition, for any trivalent vertex, the sum of the three magnetic weights 
in the groups attached to it has to be even.%
\footnote{When a pair $(p_j,q_j)$ are not relatively prime, the same set of
charges matches also those of a (reducible) product of operators. 
We find it natural to view the matching as applied to the 
irreducible representation.}

S-duality should act on these parameters generalizing the familiar exchange 
of Wilson and 't~Hooft loops in $\Ncal=4$ super Yang-Mills theory
\cite{Kapustin:2005py}. 
In the next section we identify this classification with that of 
non-self-intersecting geodesics on Riemann surfaces with hyperbolic metrics. 
These geodesics
transform in a computable way under action of the mapping class group,
providing an explicit prediction for the action of S-duality on 
these Wilson-'t~Hooft operators.

\subsection{Examples}
\label{subsec-gauge-examples}

\subsubsection*{I. $\Ncal=4$/$\Ncal=2^*$ super Yang-Mills with $G=SU(2)$}

The quiver in Figure~\ref{fig:SU(2)quivers}(a) and the torus with
one puncture in   Figure~\ref{surfaces}(a) 
represent the $\Ncal=4$ super Yang Mills theory with gauge group 
$SU(2)$, and an extra decoupled hypermultiplet. 
Alternatively, with the inclusion of a mass term for an 
adjoint hypermultiplet, this quiver represents the $\Ncal=2^*$ theory. 
Generalized Wilson-'t~Hooft loops are labeled by three integers
$(p_1,p_2;q_1)$, where
ordinary loop operators 
correspond to the case $p_2=0$.

This theory is expected to have the same S-duality symmetry as the 
$\Ncal=4$ theory.   Under S-duality  the gauge coupling 
$\tau=\theta/2\pi+4\pi i/g^2$ transforms as
\ba
\tau \mapsto \f{a \tau+b}{c\tau+d}\,,
\qquad
\begin{pmatrix}
a&b\\
c&d\\
\end{pmatrix}
\in SL(2,\Z)\,.
\ea
When the non-dynamical gauge field is turned off,
$p_2$ vanishes and the loop operators transform, 
according to \cite{Kapustin:2005py}, as
\ba
(p,q)\mapsto
(p,q)
 \begin{pmatrix}
d&-b\\
-c&a
\end{pmatrix}\,.
\label{pq-N=4}
\ea

In Section~\ref{subsec-geodesics-examples} we give an 
explicit prediction for the transformation rules 
of the loop operators also with the non-dynamical gauge fields.

\subsubsection*{II. One $SU(2)$ group with $N_F=4$}
This is the case considered in Section~\ref{sec-NF=4}.  The genus is 
$g=0$ and the number of punctures is $n=4$ so 
there is a single edge
 and two trivalent vertices, each with two open legs. 
See Figures~\ref{sphere-4pts}(a) and (b).  In the discussion 
there we did not include the non-dynamical fields, so there was a single 
magnetic weight $p_1\equiv p$ which had to be even. 

If we include the non-dynamical gauge fields $\hat A^{(2)}$, $\hat A^{(3)}$ coupling 
to the first trivalent vertex and $\hat A^{(4)}$ and $\hat A^{(5)}$ to the other one, 
then we have four more positive integers $p_2,\dots,p_5$. Now $p_1$ 
may be odd, but both $p_1+p_2+p_3$ and $p_1+p_4+p_5$ should 
be even. There is still only one electric weight $q_1\equiv q$, which is an arbitrary 
integer (unless $p_1=0$ in which case $q_1\geq0$).

It is interesting to consider S-duality in this theory, which
exchanges the r\^oles of $\gamma$ and $\delta$ in Figure~\ref{sphere-4pts}(b)
and rotates the quiver diagram in Figure~\ref{sphere-4pts}(a)
by 90 degrees \cite{Gaiotto:2009we}.
The S-duality group, which is isomorphic to $SL(2,\Z)$, 
acts on the gauge coupling 
$\tau=\theta/2\pi+4\pi i/g^2$ as
\ba
\tau \mapsto \f{a\tau+b/2}{2c\tau+d}\,,
\qquad
a,b,c,d\in\Z\,,
\qquad
ad-bc=1\,.
\ea
The transformation $\theta\ra\theta+\pi$ is a duality of this theory because
amplitudes with odd instanton numbers vanish due to fermionic zero-modes.
This is consistent with the fact that the Riemann surface is invariant
under the twist by angle $\pi$ along the geodesic $\gamma$.

As for the transformations of the loop operators, in the case where the 
non-dynamical gauge fields are turned off, this problem was
studied in \cite{Kapustin:2006hi}.
Based on the transformation of electric and magnetic charges
carried by BPS particles that can be inferred
from the Seiberg-Witten curve, it was argued that
the S-duality group acts on 
the Wilson-'t~Hooft operators by right multiplication by the inverse matrix
\ba
(p,q) \mapsto (p,q)
\begin{pmatrix}
d&-b/2\\
-2c&a
\end{pmatrix}.
\ea
This in particular implies that under the generators
\ba
S=
\begin{pmatrix}
0&-1/2\\
2&0\\
\end{pmatrix}
\qquad
\text{and}
\qquad
T=
\begin{pmatrix}
1&1/2\\
0&1\\
\end{pmatrix},
\ea
the weights transform as
\ba
S:~(p,q)\ra (-2q, p/2)\,,
\qquad
T:~(p,q) \ra (p,q-p/2)\,.
\label{ST-NF4}
\ea
The action of $T$ is the Witten effect for loop operators in this theory,
which shifts the electric weight by a multiple of the magnetic weight
under the shift of the $\theta$ angle \cite{Witten:1979ey,tHooft:1981ht}.
We will compare the transformations of weights
with the transformations of geodesics in 
Section~\ref{subsec-geodesics-examples}. 
The transformation rules of open curves provide a prediction 
for the action of S-duality on the operators with non-dynamical 
gauge fields.

\subsubsection*{III. Quiver theories for genus two surface}

Let us consider gauge theories with gauge group $SU(2)^3$ and 
no global symmetries. They can be represented as quiver diagrams 
with two vertices and three edges.  
Two examples of such theories are shown in 
Figure~\ref{fig:SU(2)quivers}(b). In the top quiver the matter fields 
transform in the fundamental of the gauge group represented by the 
central edge and as the bi-fundamental of one of the other two gauge 
groups. In the bottom quiver all matter fields are charged under all 
three gauge groups. 
The two theories are characterized by a single 
Riemann surface of genus two with no punctures shown in 
Figure~\ref{surfaces}, and thus should be related by S-duality.

In the first case, the most general supersymmetric loop operator is given by 
the set of integers
\ba
(p_j;q_j)\equiv (p_1,p_2,p_3;
q_1,q_2,q_3)
\ea
up to the usual identification under the Weyl group.
Taking the index 1 to refer to the edge in the middle,
we find the condition 
\ba
p_1+2p_2\in2\Z\,,\qquad
p_1+2p_3\in2\Z\,,\qquad
\Leftrightarrow\qquad
p_1\in2\Z\,.
\ea

In the second theory, the six integers $(p'_j;q'_j)$
that label loop operators are subject to the
condition
\ba
p_1'+p_2'+p_3'\in2\Z\,.
\ea

Apart for the expectation that S-duality is the same as maps of the 
genus 2 surface, nothing more is known about S-duality for these 
theories. Our analysis in the next section provides an 
explicit conjecture for the S-duality action on the Wilson-'t~Hooft 
operators.

\section{Curves on punctured Riemann surfaces}
\label{sec-geodesics}

As mentioned in the introduction and the preceding section, each of 
the $\Ncal=2$ theories we discuss is related to a punctured Riemann surface. 
From the M-theory descriptions of loop operators in these theories we are 
motivated to look at curves on this Riemann surface. 
A loop operator will arise from an M2-brane 
ending on a collection of M5-branes along a two dimensional manifold 
with one direction in space-time and the other on the Riemann 
surface. In Appendix~\ref{app-ads} we study the supergravity duals 
of the theories based on $SU(N)$ at large $N$, where the curves turn out 
to be geodesics with respect to the hyperbolic metric on the 
Riemann surface.

From the gauge theory side we saw that 
the quiver in Figure~\ref{sphere-4pts} corresponds to a specific 
representation of the Riemann surface in terms of ``pairs of pants'' 
(see below). In fact a theorem due to Dehn classifies 
homotopy classes of non-self-intersecting curves on Riemann surfaces 
using exactly this representation of the Riemann surface. As 
we shall see, this classification matches perfectly that of the loop 
operators in the previous section.

We begin by describing the topology of an oriented punctured Riemann surface
and the Fenchel--Nielsen coordinates for Teichm\"uller space,
following \cite{MR2249478} and \cite{MR963064}.
For any (oriented) punctured Riemann surface $\Sigma$ of negative Euler
characteristic with genus $g$ and $n$
punctures, there exist $3g-3+n$  pairwise disjoint connected curves
without self-intersection, none of which is homotopic to zero or
to a curve arbitrarily close to a puncture, such that the complement of
the curves is a union of $2g-2+n$ pairs-of-pants.  We treat a puncture
as a pants-leg of zero length, so some of these pairs-of-pants may look
like a punctured annulus or a twice-punctured disk.  

\begin{figure}[ht]
\begin{center}
\includegraphics[scale=.4]{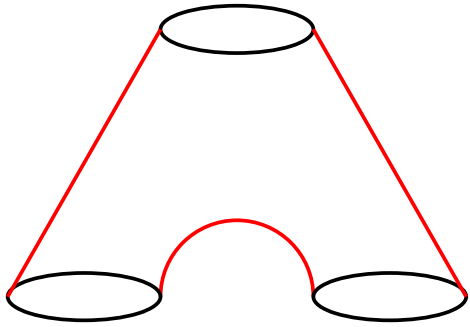}
\end{center}
\caption{A pair of pants with seams}\label{fig:pairofpants}
\end{figure}

\begin{figure}[ht]
\begin{center}
\includegraphics[scale=.3]{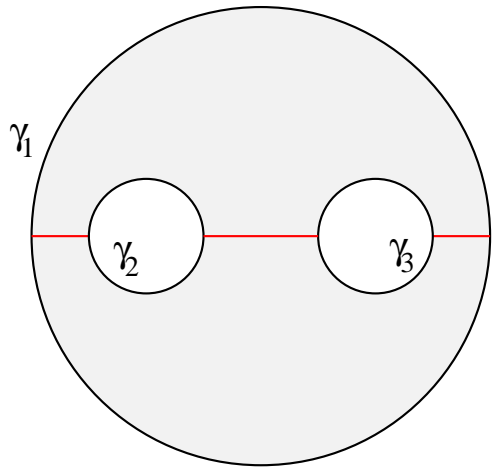}
\end{center}
\caption{A pair of pants, shown as a disk with two holes}\label{fig:pants+seams}
\end{figure}

This decomposition is topological, but if a hyperbolic metric has been
specified on $\Sigma$, then the curves giving the decomposition can be
chosen to be geodesics (after a homotopy), and the resulting pairs of
pants inherit constant curvature metrics.  The lengths of the separating
geodesics are determined by the metric; on the other hand, once the
lengths of the three pants-legs have been specified, a
 pair-of-pants has a unique metric of constant curvature $-1$, and there
is a unique geodesic arc (the ``seam'') connecting each pair of pants-legs,
as shown in Figure~\ref{fig:pairofpants}.  An alternate visualization
is given in Figure~\ref{fig:pants+seams}

To re-assemble such a collection of pairs-of-pants into a punctured Riemann
surface with metric
(assuming that we have fixed in advance the set of pants-legs which have
length zero and correspond to punctures),
we need to specify both the length $\ell_j\in\mathbb R^+$
and a twist parameter $\theta_j\in \mathbb R$ for each pants-leg of
non-zero length.  The twist parameter measures the relative angular twist when
gluing together the two pairs-of-pants along a common pants-leg, {\it i.e.,}
the displacement of the seams.  (Note
that although
twisting by an integer multiple of $2\pi$ gives a diffeomorphic surface,
the diffeomorphism is not isotopic to the identity; this diffeomorphism
is known as a {\em Dehn twist}.)  Fenchel and Nielsen
showed that this assignment establishes an isomorphism between the associated
Teichm\"uller space and $(\mathbb R^+)^{3g-3+n}\times \mathbb R^{3g-3+n}$.

There are many possible choices of pants decomposition (and of the
corresponding Fenchel--Nielsen coordinates); we
fix one such and
use it to describe all homotopy classes of closed curves on $\Sigma$
without self-intersection, as well as homotopy classes of
arcs connecting punctures  without self-intersection.
The result is known as
Dehn's theorem.\footnote{This theorem was found by
Dehn in the 1920's (but only published much later \cite{dehn-breslau}), 
and rediscovered by Thurston in the 1970's (also published much 
later \cite{MR956596}).
We follow the account in
\cite{MR1144770}. }
We do not assume that the curve is connected; we {\em do}\/ assume that no
component of the curve is homotopic to zero, and that no component
of the curve is homotopic to a curve arbitrarily close to one of
the punctures.

If a hyperbolic metric has been specified on the Riemann surface, then
each homotopy class of curves or arcs without self-intersection contains
a unique geodesic (which also is without self-intersection).
Dehn's theorem can therefore be viewed as a classification of
non-self-intersecting geodesics, rather than a classification of homotopy
classes of non-self-intersecting curves.

\begin{figure}[ht]
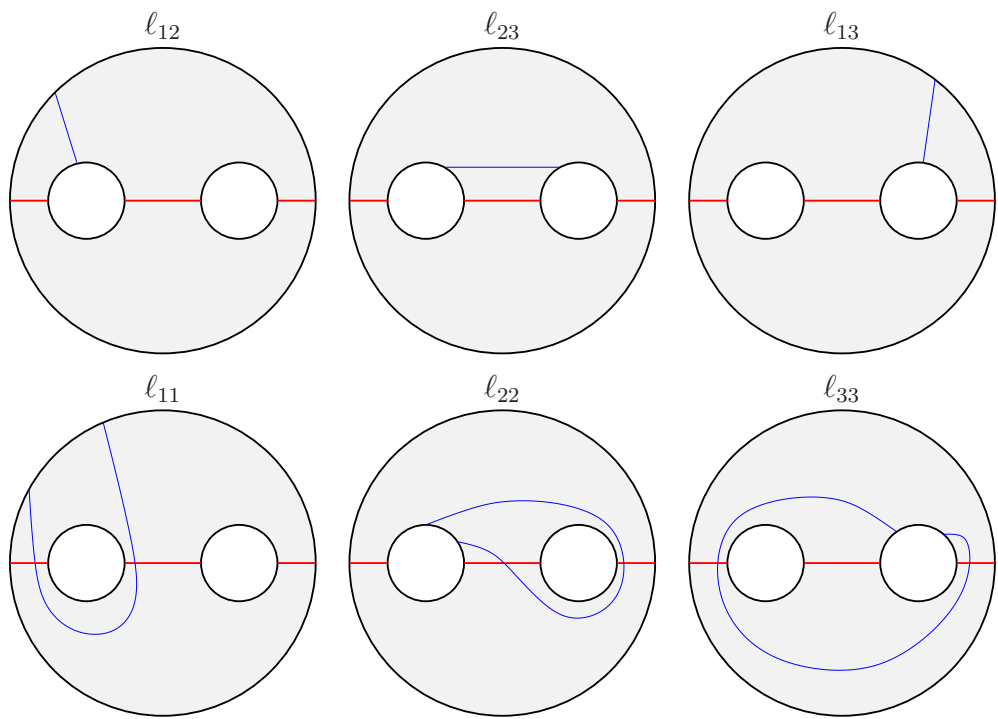

\begin{tabular}{ccc}
$\ell_{12}$&$\ell_{23}$&$\ell_{13}$\\
\includegraphics[scale=.2]{l12.mps}&
\includegraphics[scale=.2]{l23.mps}&
\includegraphics[scale=.2]{l13.mps}\\
$\ell_{11}$&$\ell_{22}$&$\ell_{33}$\\
\includegraphics[scale=.2]{l11.mps}&
\includegraphics[scale=.2]{l22.mps}&
\includegraphics[scale=.2]{l33.mps}\\
\end{tabular}
\caption{Basic arcs on a pair-of-pants}\label{fig:arcs}
\end{figure}

The first step in Dehn's analysis is to characterize collections of arcs without
self-intersection on a single pair-of-pants.  Dehn showed that any such
collection of arcs can be moved, by a homotopy that leaves the
endpoints of the arcs on the boundary, to a collection of arcs
whose boundary points lie in the upper semicircles of each of
the boundaries, and each of which is homotopic to one
of six specific ``basic'' arcs, illustrated in Figure~\ref{fig:arcs}.

Such a collection of arcs determines three non-negative
integers
$p_1$, $p_2$, and $p_3$ which tell how many endpoints there are
on each boundary circle $\gamma_1$, $\gamma_2$, and $\gamma_3$,
subject to the condition that $p_1+p_2+p_3$ is even.

Conversely, given any three non-negative integers $p_1$, $p_2$, and
$p_3$ whose sum is even, there is a collection of non-intersecting
arcs of the six
basic types, unique up to homotopy, with the desired numbers of endpoints.
To see this, first note that if $p_i > p_j+p_k$ and $p_j> p_i+p_k$ then
$p_i+p_j > p_i + p_j + 2p_k$ so that $p_k$ is negative, a contradiction.
Thus, at most one of the numbers $p_i-p_j-p_k$ is positive.

If $p_i > p_j+p_k$, then we can use
\be
\frac12(p_i-p_j-p_k)\ell_{ii} + p_j \ell_{ij}+p_k \ell_{ik}.
\ee
On the other hand, if $p_j+p_k\ge p_i$ for all permutations of $\{1,2,3\}$,
then we can use
\be
 \frac12(p_1+p_2-p_3)\ell_{12} + \frac12(p_1+p_3-p_2)\ell_{13} +
\frac12(p_2+p_3-p_1)\ell_{23}.
\ee

\begin{figure}[ht]
\begin{center}
\includegraphics[scale=.3]{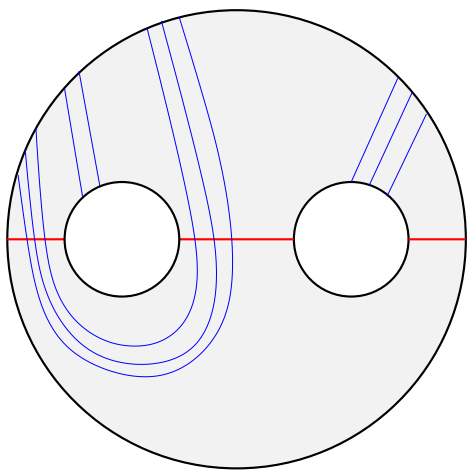}
\end{center}
\caption{Example of arcs: $p_1=11$, $p_2=2$, $p_3=3$}\label{fig:arc}
\end{figure}

The second step in Dehn's theorem is to attach the endpoints on either
side of a boundary circle. 
One possibility is that the given non-self-intersecting
curve has intersection number $0$
with $\gamma_j$ so that no attachment needs to be done; 
in this case, we define the {\em twisting number}\/ 
$q_j$ to be the number of
components of the curve which are homotopic to $\gamma_j$. 
(Thus, $p_j=0$ implies $q_j\ge0$.)

\begin{figure}[ht]
\begin{center}
\includegraphics[scale=.4]{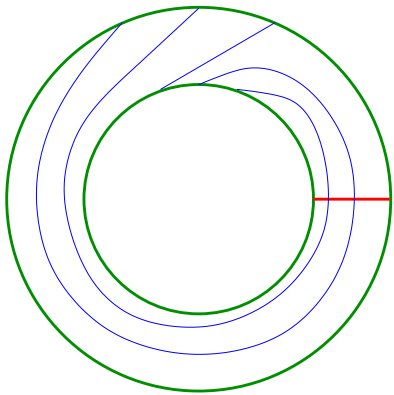}
\end{center}
\caption{Example of twisting: intersection number $p=3$,
 twisting number $q=+2$.}\label{fig:twisting}
\end{figure}

Otherwise, when $p_j\neq0$, there is a canonical gluing of the curves 
along the boundary $\gamma_j$, since 
using the basic arcs in Figure~\ref{fig:arcs}, 
the $p$ endpoints are on one half of the boundary circle. 
This gluing is the one labeled by $q_j=0$. As mentioned, it is 
possible to twist the curves before gluing, which means that this 
identification will not be obeyed. After a homotopy, the twisting can
be confined to a small annulus containing the curve, as illustrated
in Figure~\ref{fig:twisting}.
The twisting number $q_j$ is positive if the attaching curves bend to
the right (as in the Figure), and is negative if the attaching curves
bend to the left.  The magnitude $|q_j|$ of the twisting number
counts the number of shifts when matching the arcs on the two sides.
A simple way to calculate $|q_j|$ is to count the number of intersections
of the curve with a fixed arc which connects the two boundaries of the
annulus (as illustrated in Figure ~\ref{fig:twisting}).
Note that if we make a Dehn twist along $\gamma_j$ (in the 
direction which curves to the right), 
the twisting number changes by $q_j\mapsto q_j+p_j$.

The intersection number  $p_j$ and the twisting number
$q_j$ give a complete local description of how the given curve behaves
near $\gamma_j$.  For the local behavior near
each puncture, we specify a corresponding intersection number $p_j$ (the
number of endpoints at the puncture, or the intersection with a small
loop near the puncture) but do not need to specify a twisting number $q_j$,
since there is no winding at the puncture (up to homotopy).

The precise statement of Dehn's theorem involves intersection numbers
between non-self-intersecting curves.  Since the topological version of
the classification involves homotopy classes of curves rather than specific
curves, it is natural to define $\#(\gamma\cap\delta)$ to be the minimum
of the number of intersection points as $\gamma$ and $\delta$ vary among
non-self-intersecting curves in their respective homology classes.  (Note
that all intersections are counted {\em positively}---there is no sign or
orientation taken into account.)  This definition becomes much simpler
when $\Sigma$ has been given a hyperbolic metric: it turns out that one
geodesic from each of the two homotopy classes can be used to compute 
the intersection
number $\#(\gamma\cap\delta)$ directly, just by counting intersection points
of the geodesics.

\bigskip

\noindent {\bf Dehn's Theorem.}
Let $\Sigma$ be an oriented punctured Riemann surface  of negative Euler
characteristic with genus $g$ and $n$
punctures.  Let $\gamma_1$, \dots, $\gamma_{3g-3+n}$ be 
 pairwise disjoint connected curves without self-intersection
whose complement is a pants decomposition of $\Sigma$, and let 
$\gamma_{3g-3+n+1}$, \dots, $\gamma_{3g-3+2n}$
be simple closed curves 
near the punctures.  Define a
mapping
\be
\gamma \mapsto (\#(\gamma\cap \gamma_j); q_j) \in
(\mathbb Z_{{}\ge0})^{3g-3+2n}\times \mathbb Z^{3g-3+n}
\ee
which assigns to each homotopy class of closed curves
 without self-intersection or
arcs connecting punctures without self-intersection its intersection
number $p_j=\#(\gamma\cap \gamma_j)$
with $\gamma_j$, $1\le j\le 3g-3+2n$
 and its twisting number $q_j$ with respect to $\gamma_j$, $1\le j\le 3g-3+n$.
(Note that these intersection and twisting numbers depend only on the
homotopy class of $\gamma$.)
Then this mapping is one-to-one, and its image is
\begin{align*} \{ (p_1,p_2,\ldots,&\, p_{3g-3+2n};q_1,q_2,\ldots, q_{3g-3+n}) 
\\
& |\ \text{ if } p_j=0 \text{ then } q_j\ge0,
\text{ and } p_{i}+p_{j}+p_{k} \in 2\mathbb Z
\\ &\text{ whenever } \gamma_{i} \cup \gamma_{j} \cup \gamma_{k}
\text{ is the boundary of a pair-of-pants} \}.
\end{align*}
The integers $p_j$, $q_j$ are called the {\em Dehn--Thurston parameters}\/ of
$\gamma$.

\bigskip

It is easy to see that Dehn's classification of homotopy classes of
non-self-intersecting 
curves (or of non-self-intersecting geodesics when a hyperbolic metric
is being used)  is in one-to-one correspondence with the classification 
of Wilson-'t~Hooft loop operators in the gauge theory 
in Section~\ref{sec-CFT}, see equation (\ref{gen-weights2b}). The twisting
numbers $q_j$ match the electric charges carried by 
Wilson loops and the intersection numbers
$p_j$ match the magnetic charges charges carried by 
't~Hooft loops. 
Moreover, the identification respects the expected Witten effect on loop
operators \cite{Witten:1979ey,tHooft:1981ht}:  when the $j^{\text{th}}$
theta angle of the gauge theory is
increased by $2\pi$, corresponding to performing a Dehn twist 
of $\Sigma$ along $\gamma_j$, the Dehn--Thurston parameters change
by $(p_j,q_j)\mapsto (p_j,q_j+p_j)$.
This identification also matches the intuition one 
gets from the M-theory constructions, both from looking at M2-branes 
ending on M5-branes and from the M2-branes in the supergravity 
background as discussed in Appendix~\ref{app-ads}.

The formulation of Dehn's theorem, as stated, depends on the choice of
a pants decomposition (or on the corresponding
choice of Fenchel--Nielsen
coordinates).  It is natural to ask how the data describing 
non-self-intersecting geodesics changes when the pants decomposition
changes.
This is essentially the question of S-duality
for the field theory: according to Gaiotto's classification 
\cite{Gaiotto:2009we}, each choice of pants decomposition
corresponds to a choice of S-duality frame of the 
field theory. In geometric terms, these S-dualities
are given by the mapping class
group of the surface.

The {\em mapping class group}\/  is the familiar discrete group by which
Teichm\"uller space must be quotiented in order to get the moduli space
$\mathcal{M}_{g,n}$ of Riemann surfaces of genus $g$ with $n$ marked points;
this group relates the various choices of pants decomposition to each other.
It is known \cite{MR0171269} that the mapping class group is generated
by Dehn twists on a small set of geodesic loops.  In our setup, some of
these will be Dehn twists along the boundaries between pairs of pants
(which are easy to interpret in the field theory, as shifts in the theta
angles).  Others, though, are Dehn twists along loops which cross those
boundaries, and these are more difficult to analyze.

Penner \cite{MR743669} gave a general description of the action of the
mapping class group on the Dehn--Thurston parameters\footnote{Penner's
theorem was explicitly formulated and proven for the case with no punctures.
But as he remarks at the end of his paper, the theorem
 remains true when punctures
are allowed, and it is the version with punctures which we have stated here.}
$(\mathbb Z_{{}\ge0})^{3g-3+2n}\times \mathbb Z^{3g-3+n}$ as follows:
for each element $\varphi$ of the mapping class group,
there is a decomposition $K_\varphi$
of the vector space $\mathbb R^{6g-6+3n}$ into
a finite number of cones based at the origin such that on each cone in
$K_\varphi$, $\varphi$ acts like an invertible integer matrix.
(Following Thurston \cite{MR956596}, Penner calls these
 piecewise-integer-linear
(or PIL) transformations.)

\begin{figure}[ht]
\begin{center}
\psfrag{gamma}{$\gamma$}
\psfrag{delta}{$\delta$}
\includegraphics[scale=.3]{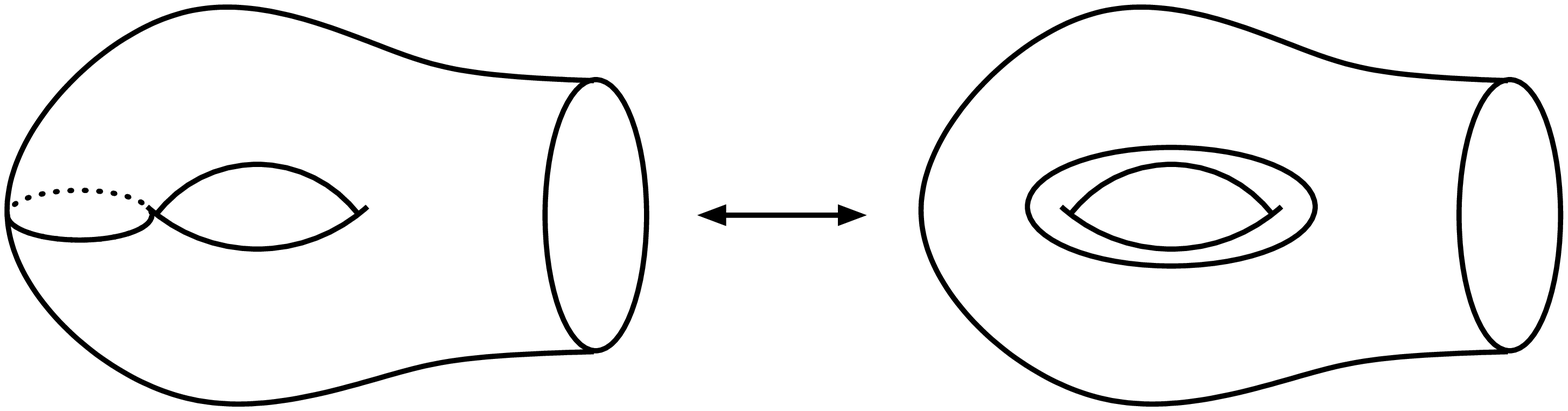}
\\
\includegraphics[scale=.3]{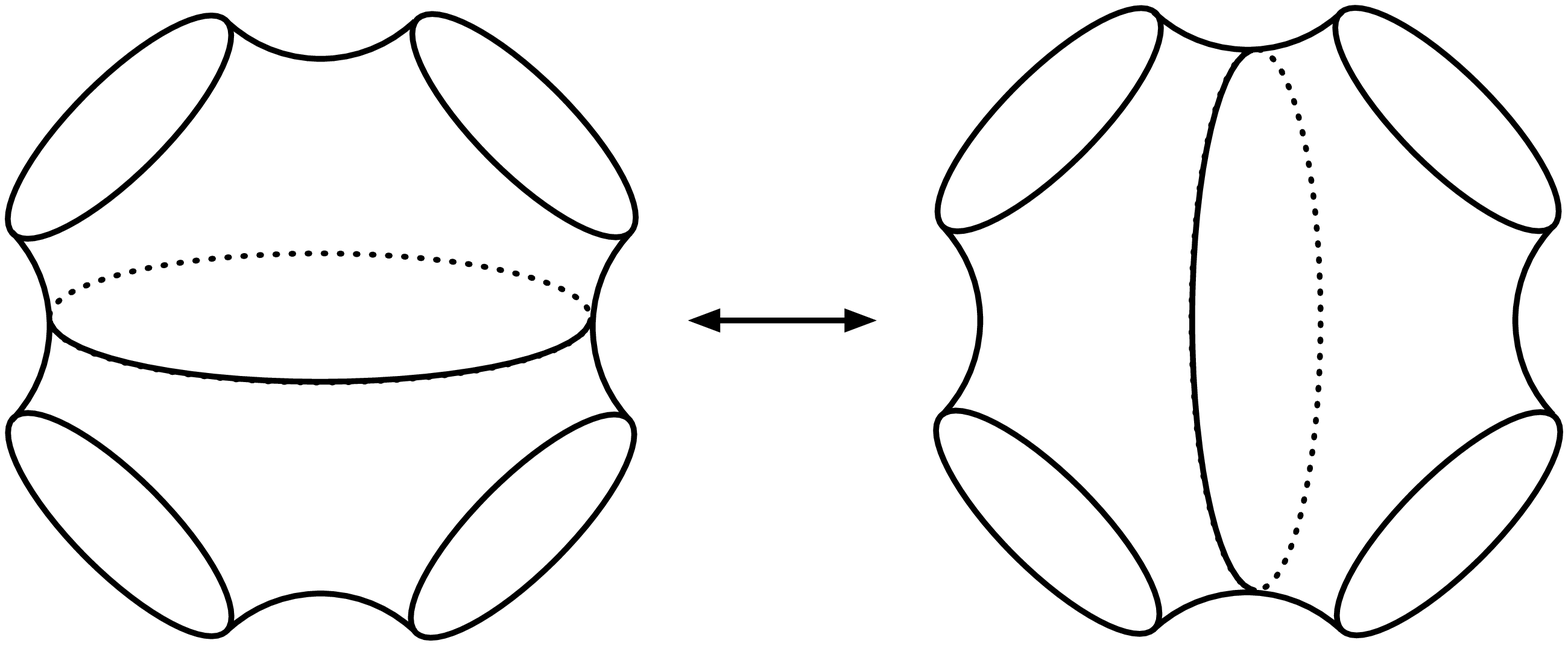}
\parbox{5in}{
\caption{
Two types of elementary transformation.
\label{hatcher-thurston}}}
\end{center}
\end{figure}

Due to a result of Hatcher and Thurston \cite{MR579573}, any two
pants decompositions of $\Sigma$ can be obtained from one another
by a sequence of ``elementary transformations'' of two types,
illustrated in Figure~\ref{hatcher-thurston}.\footnote{The original formulation
of Hatcher and Thurston did not allow punctures, but the result is now
known to extend to the punctured case: see \cite{MR2040283}, for example.}
Penner explicitly
computes the PIL transformation on the Dehn--Thurston parameters for each
of the two types of elementary transformation, 
and this is enough to determine it for
an arbitrary change of pants decomposition.  
Working out the transformation rules in specific examples is not 
too difficult, but Penner's formulas for general curves and 
transformations are extremely complicated. We refer the 
reader to the original paper
\cite{MR743669} and the book \cite{MR1144770} rather than trying to reproduce those formulas here.  However, we
will give special cases of Penner's formulas in some examples in the next
subsection.

A change in pants decomposition also induces a change in Fenchel--Nielsen
coordinates, that is, in the lengths and twists of the basic geodesics.
An explicit coordinate change formula for these has been computed by 
Okai \cite{MR1270159} (again for the two types of elementary transformation),
which may be useful in future explorations of S-duality for these theories.

\subsection{Examples: checks and predictions}
\label{subsec-geodesics-examples}

\subsubsection*{I. A torus with one puncture}

By gluing two pants-legs of a pair of pants 
together while degenerating the other to a puncture, one obtains
a torus with one puncture.
As we discussed in Section~\ref{subsec-gauge-examples},
this Riemann surface corresponds to $\Ncal=2^*$
super Yang-Mills as well as $\Ncal=4$.
All geodesics without self-intersection 
are classified by their intersection number $p_1\geq 0$
with $\gamma$,
the twisting number $q_1\in \Z$ with respect to $\gamma$,
and the number $p_2$ of end points at the puncture.

We apply the first type of Hatcher--Thurston transformation (shown in the
upper half of Figure~\ref{hatcher-thurston}) to obtain $(p_1',p_2';q_1')$
from $(p_1,p_2;q_1)$.  Since $p_2$ represents the number of end-points at
the puncture, it is always even and $p_2'=p_2$.  The formula for the
other parameters must be divided into cases, according to which type of
arc the given charges define on the pair-of-pants, both before and
after the transformation. 
This division into cases gives the piecewise-linear
structure of the action of the transformation on the charges.  The cases
are as follows:
\begin{itemize}
\item{\bf Case 1:}
If $p_2>2p_1$ and $p_1>|q_1|$ then
\begin{align*}
p_1' &= \frac{p_2}{2} - p_1 + |q_1|\\
q_1' &= -q_1
\end{align*}
\item{\bf Case 2:}
If $p_2>2p_1$ and $p_1\le|q_1|$ then
\begin{align*}
p_1' &= \frac{p_2}{2} - p_1 + |q_1|\\
q_1' &= -\sign(q_1) p_1
\end{align*}
Note that if $q_1=0$ then $p_1=0$, so the last line is well-defined.
\item{\bf Case 3:}
If $p_2\le2p_1$ and $p_2>2|q_1|$ then
\begin{align*}
p_1' &= |q_1|\\
q_1' &= -\sign(q_1)\left(p_1-\frac{p_2}{2}+|q_1|\right)
\end{align*}
where we define $\sign(0)=-1$.
\item{\bf Case 4:}
If $p_2\le2p_1$ and $p_2\le2|q_1|$ then
\begin{align*}
p_1' &= |q_1|\\
q_1' &= -\sign(q_1) p_1
\end{align*}
Note that if $q_1=0$ then $p_1=0$, so the last line is well-defined.
\end{itemize}
When $p_2=0$, we are in case 4, and the formula reproduces the prediction
from gauge theory (\ref{pq-N=4}): it is the standard
S-duality transformation from $SL(2,\mathbb Z)$, adjusted so that both
$p_1$ and $p_1'$ are non-negative. 
All other cases are new predictions 
for the action of the S-duality group on generalized loop operators with 
non-trivial bundles of the flavor groups.

The formula above is a special case of Penner's formula \cite{MR743669}, 
but can be obtained directly as follows:
the value for the new intersection number $p_1'$ is obtained by counting the number
of times the arcs cross the line segment which becomes a boundary circle after
the transformation.  The absolute value of the new twisting number $q_1'$ is then
determined by making the formula invertible.  To determine the sign
of the new twisting number, we need to analyze the diagram of how the arcs
change, which we do for each case (in a well-chosen example) below.

\begin{figure}[ht]
\begin{center}
\begin{tabular}{ccc}
\includegraphics[scale=.23]{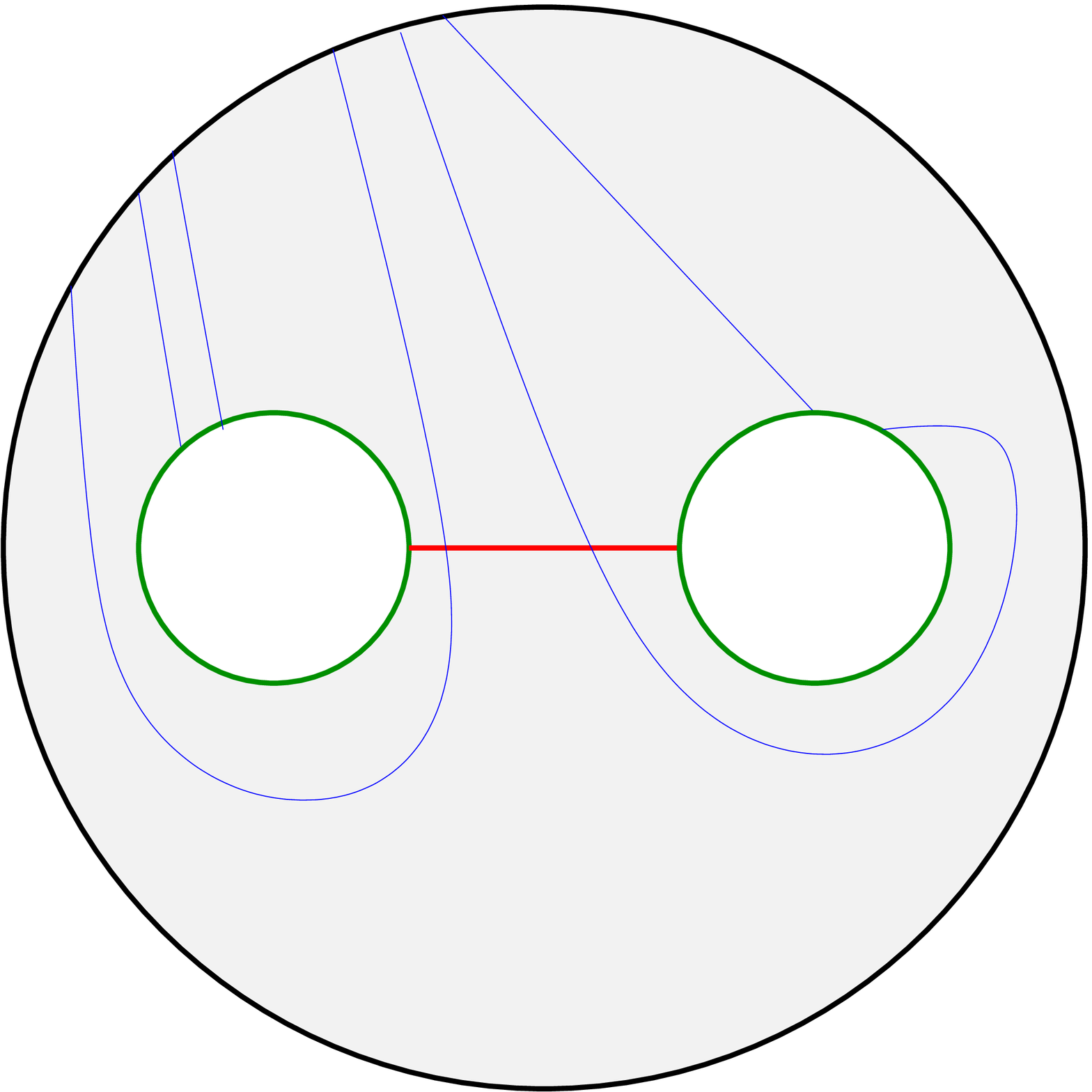}&
\includegraphics[scale=.23]{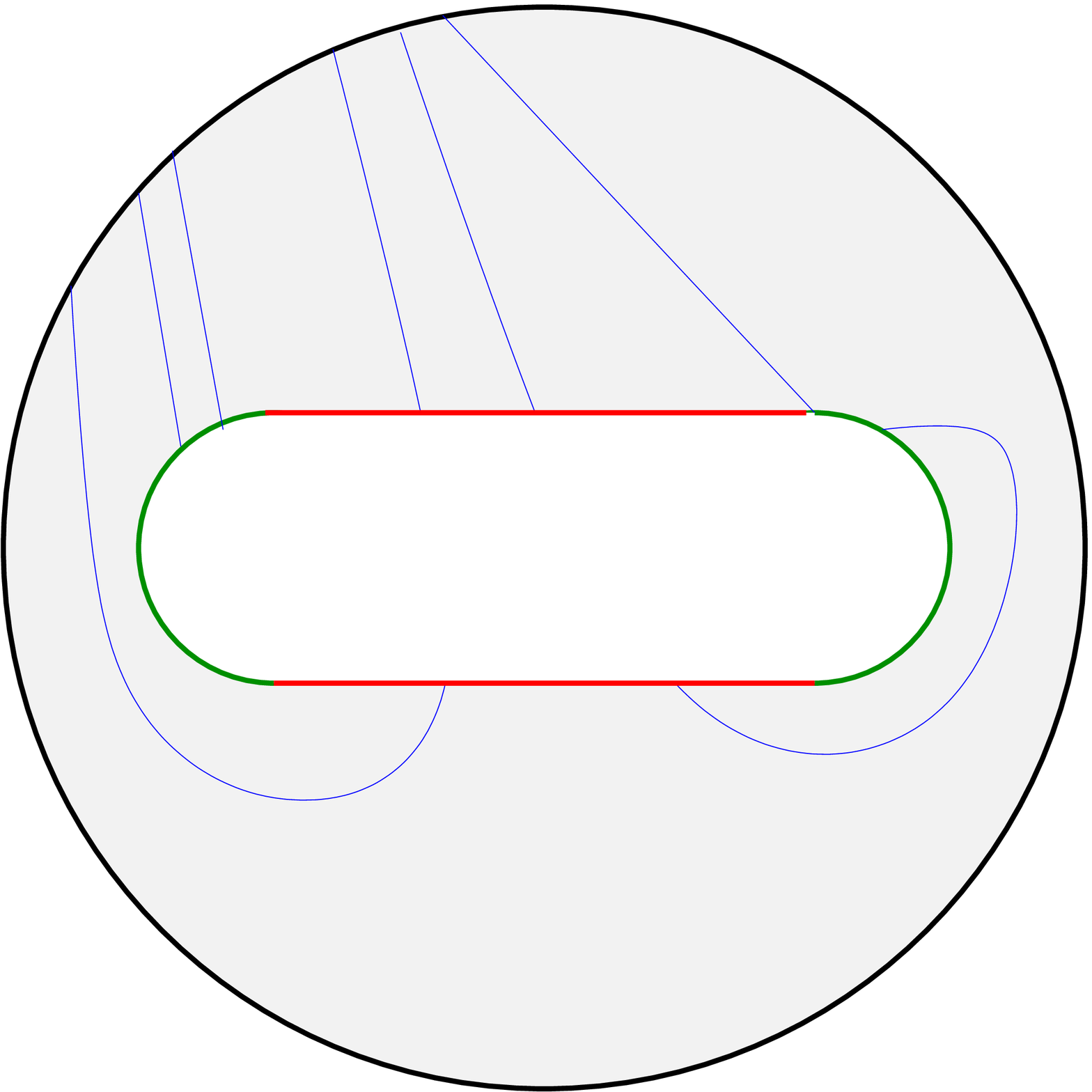}&
\includegraphics[scale=.23]{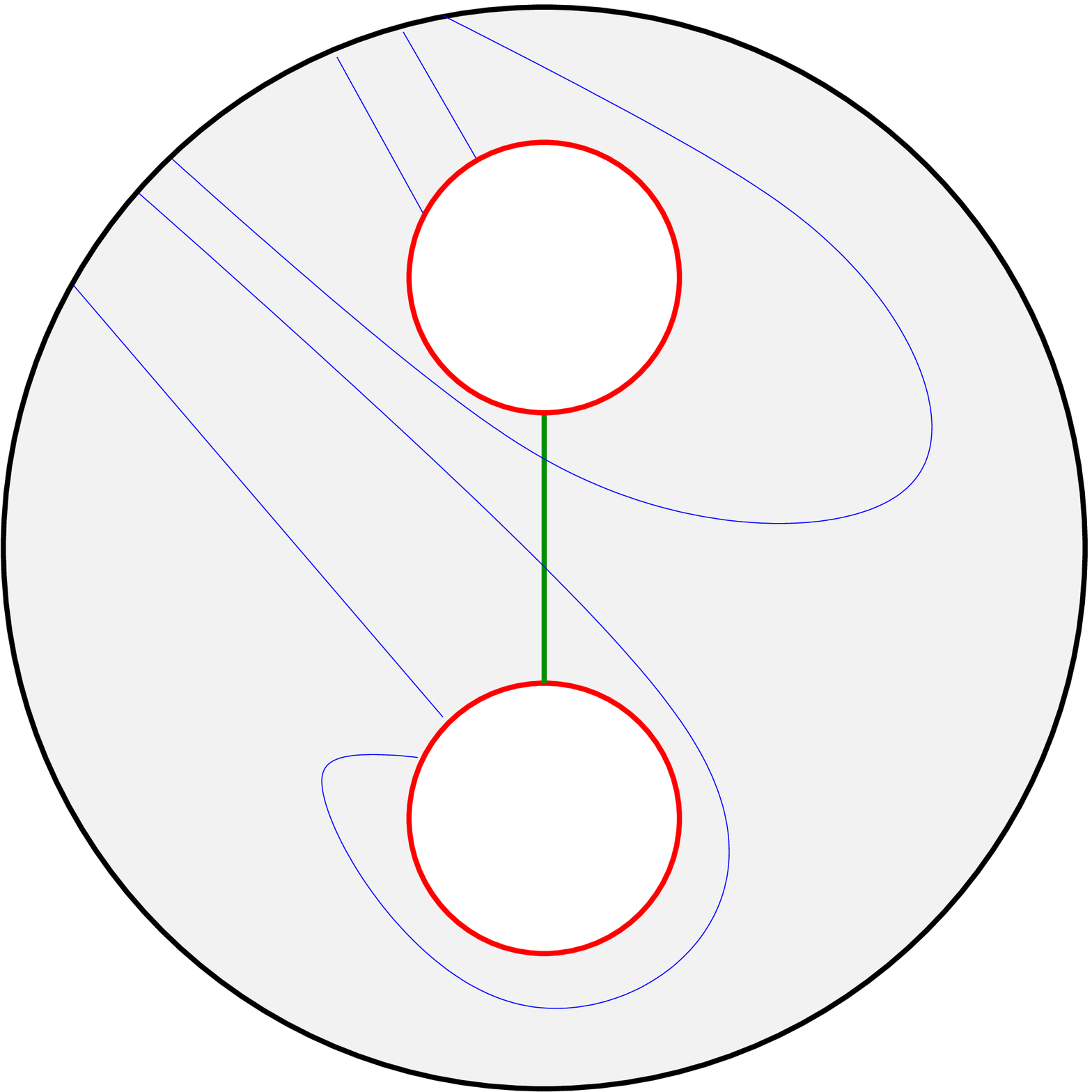}
\end{tabular}
\caption{The transformation, step by step.}\label{transform1}
\end{center}
\end{figure}

In order to explain how to apply the transformation in a particular
example, we start with 
$(p_1,p_2;q_1)=(2,6;1)$, depicted on the left side of Figure~\ref{transform1}.
We open the pair of pants up along the seam, as depicted in the
center of Figure~\ref{transform1}.
Then we re-attach the other way, leaving boundary circles in place of the 
edges
of the original seam, as depicted on the right side of Figure~\ref{transform1}.
In this way, we obtain $(p_1',p_2';q_1')=(2,6,-1)$, and we see that the
sign of $q_1'$ is correct in the above formula.

\begin{figure}[ht]
\begin{center}
\begin{tabular}{lll}
$(1,4;1)$ && $(2,4;-1)$\\
\includegraphics[scale=.2]{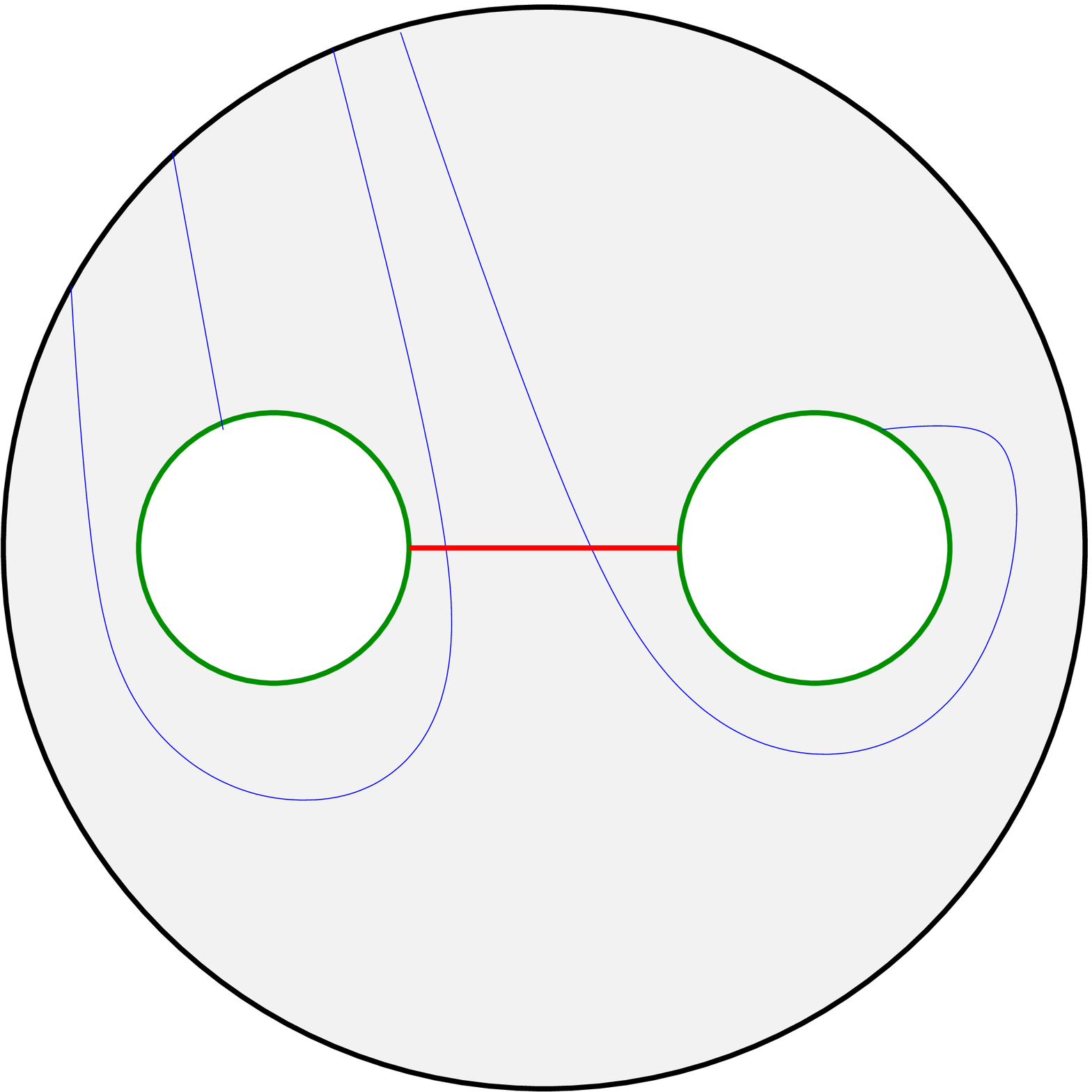}&
\qquad&
\includegraphics[scale=.2]{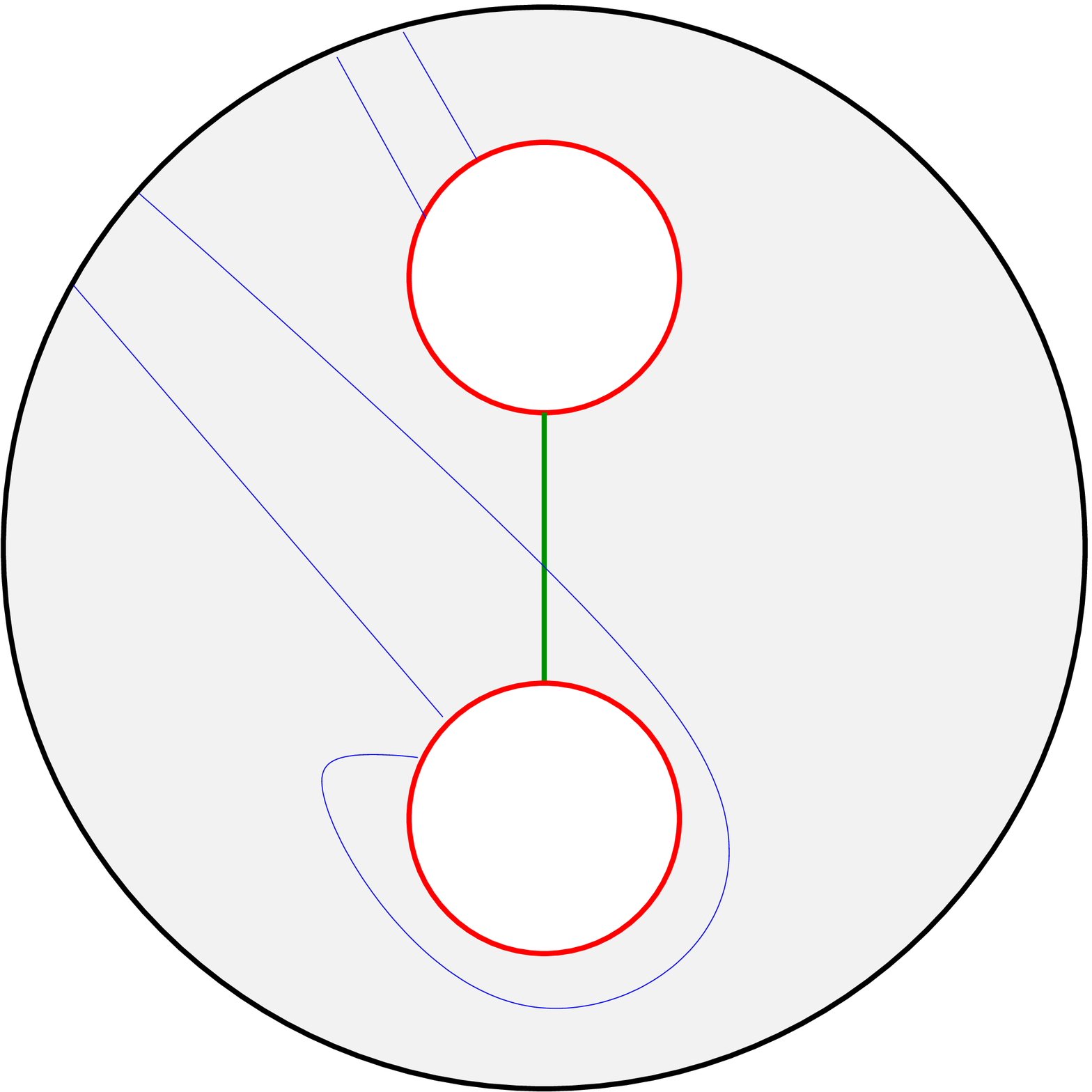}\\
$(1,2;1)$ && $(1,2;-1)$\\
\includegraphics[scale=.2]{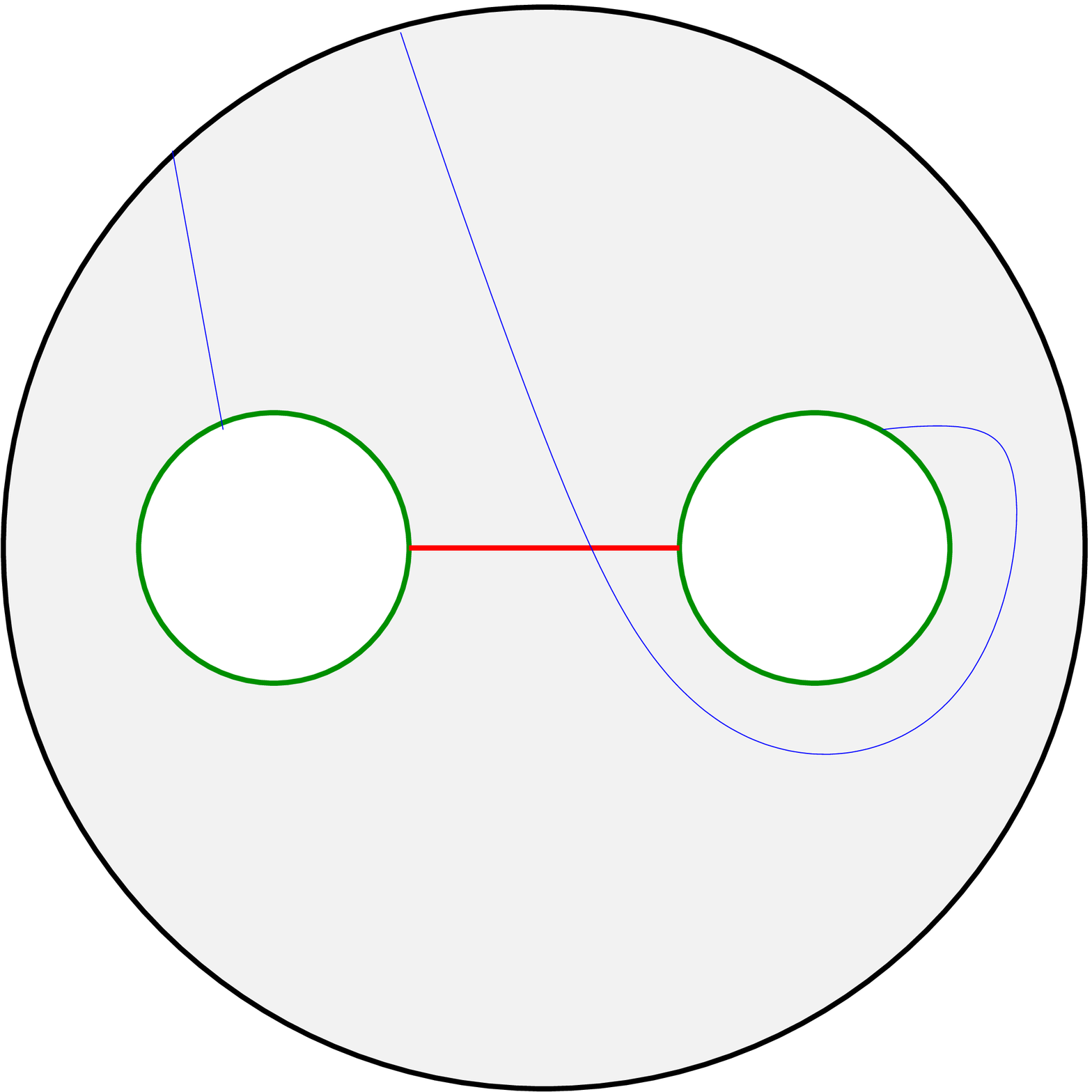}&
\qquad&
\includegraphics[scale=.2]{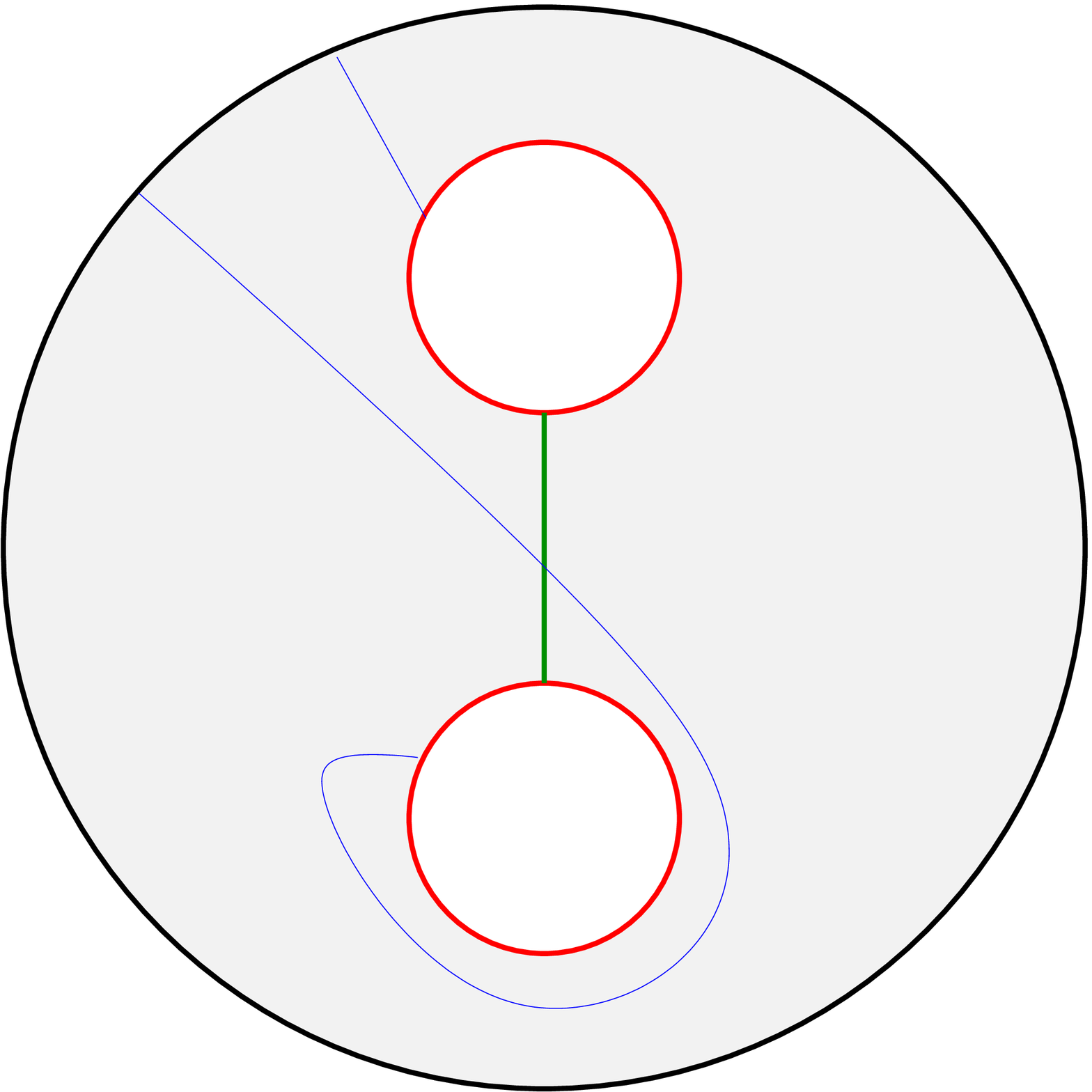}
\end{tabular}
\caption{The effect of the transformation on several arcs.}\label{transforms}
\end{center}
\end{figure}

Similarly, if we choose $(p_1,p_2;q_1)=(1,4;1)$ (depicted in the upper left
corner of Figure~\ref{transforms}) and apply the transformation, we
find the arc in the upper right corner of Figure~\ref{transforms}),
with $(p_1',p_2';q_1')=(2,4;-1)$.
The end result has $q_1'=-1$, consistent with the second case of our
formula.  The third case is simply the inverse of the second case, so 
we do not need to illustrate that one separately.

Finally, if we choose $(p_1,p_2;q_1)=(1,2;1)$ (depicted in the lower left
corner of Figure~\ref{transforms}) and apply the transformation, we
find the arc in the lower
right corner of Figure~\ref{transforms}),
with $(p_1',p_2';q_1')=(1,2;-1)$.
The end result has $q_1'=-1$, consistent with the fourth case of our formula.

\subsubsection*{II. A sphere with four-punctures}

\begin{figure}[ht]
\begin{center}
\includegraphics[scale=.35]{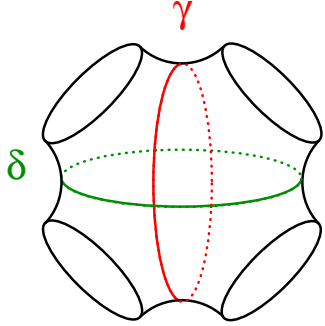}
\caption{Four-punctured sphere with two meridians.}
\label{two-meridians}
\end{center}
\end{figure}

The previous example involved the first type of elementary transformation,
but we must now consider the second type, whose action on the
Dehn--Thurston parameters is much more complicated.  
If we restrict to closed geodesics, 
we only need a very special case of the transformation,
but in the next example we will need to grapple with the complications.

We begin with the four-punctured sphere, illustrated 
in Figure~\ref{two-meridians} showing the two meridians $\gamma$, $\delta$
which represent two possible decompositions
into pairs of pants. (We also enlarged the punctures into disks.)  
A closed non-self-intersecting geodesic
 with parameters $(p,q)$ (with respect to
$\gamma$) will meet $\gamma$ $p$ times, and wind around the surface $q$
times parallel to $\gamma$.  During each full revolution of winding,
the geodesic will meet the other meridian $\delta$ two times.  Thus, 
the total intersection with $\delta$ is $2|q|$.  In other words,
\be
\begin{aligned}
\#(C\cap\gamma) &= p\\
\#(C\cap\delta) &= 2|q|.
\end{aligned}
\ee
Applying the same analysis to the meridian $\delta$ (in which the 
r\^oles of $\gamma$ and $\delta$ are reversed), we find
\be
\begin{aligned}
p' = \#(C\cap\delta) &= 2|q|\\
|q'| = \frac12\#(C\cap\gamma) &= \frac{p}2.
\end{aligned}
\ee
By carefully considering the orientation of winding, one can
show that the signs of $q$ and $q'$ are opposite.  Thus, the S-duality
transformation is
\be
(p,q) \mapsto \left(2|q|,-\sign(q)\frac{p}2\right),
\ee
verifying equation 
(\ref{ST-NF4}) from the gauge theory (provided that we use
the Weyl group to make the magnetic charge positive).

In the same way one can analyze also open geodesics with 
endpoints at the punctures. This would provide a prediction for the action 
of S-duality on the loop operators with non-dynamical gauge fields. 
While this is rather simple to do in specific examples (see also the next 
subsection), the general case is rather messy to analyze.

\subsubsection*{III. Genus 2 }

When the Riemann surface $\Sigma$ has genus two and no punctures, there
are two possible decompositions into pairs of pants, represented
by the quivers shown in
Figure~\ref{fig:SU(2)quivers}(b).  Both types of Hatcher--Thurston
transformation can act here: in the case of the upper quiver from
Figure~\ref{fig:SU(2)quivers}(b), either
of the two pairs of pants in the decomposition is a torus with a disk
removed, and the first type of Hatcher--Thurston transformation can be
applied.  We gave a fairly complete formula for this in example I above,
omitting only the second twist parameter (which corresponds to the
twist around the loop connecting the two pieces into $\Sigma$).
The complete formula can be found in 
the references \cite{MR1144770,MR743669}.

\begin{figure}[ht]
\begin{center}
\includegraphics[scale=.35]{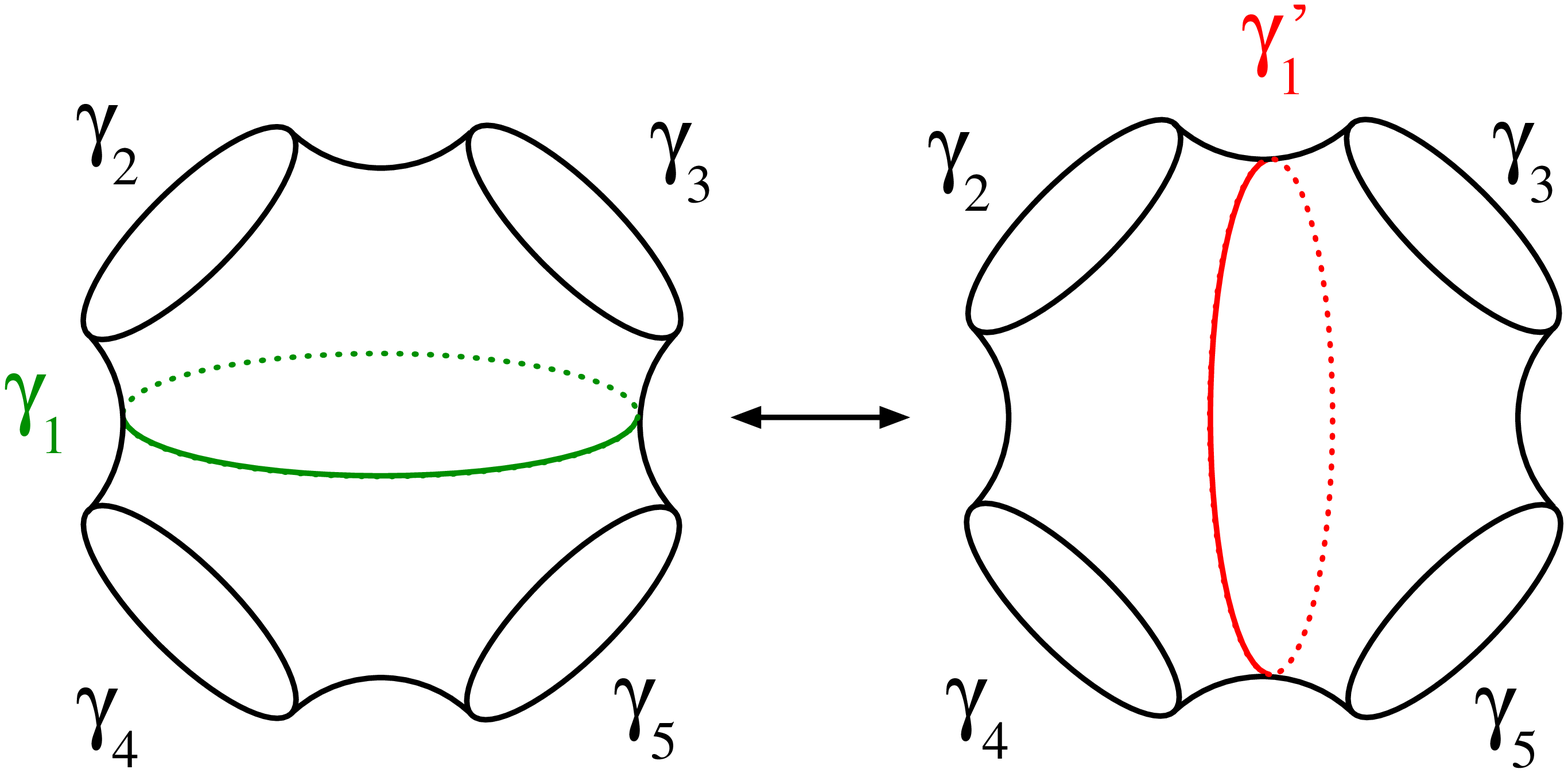}
\caption{The second type of elementary transformation.}
\label{hatcher-thurston-labeled}
\end{center}
\end{figure}

The second type of Hatcher--Thurston transformation can also act,
exchanging the two quivers in Figure~\ref{fig:SU(2)quivers}(b).
To describe the action of this transformation on the Dehn--Thurston
parameters, we illustrate the transformation again
in Figure~\ref{hatcher-thurston-labeled}, where we
have labeled the boundary circles $\gamma_2$, \dots, $\gamma_5$ as well
as the meridians $\gamma_1$ and $\gamma_1'$ which are used to decompose
the surface into two pairs-of-pants, before and after the transformation.
For our application to the case of genus two, we identify $\gamma_4$ with
$\gamma_2$, and $\gamma_5$ with $\gamma_3$.  Thus, the Hatcher--Thurston
transformation moves us from the lower quiver in
Figure~\ref{fig:SU(2)quivers}(b) to the upper quiver.

\begin{figure}[ht]
\begin{center}
\includegraphics[scale=.27]{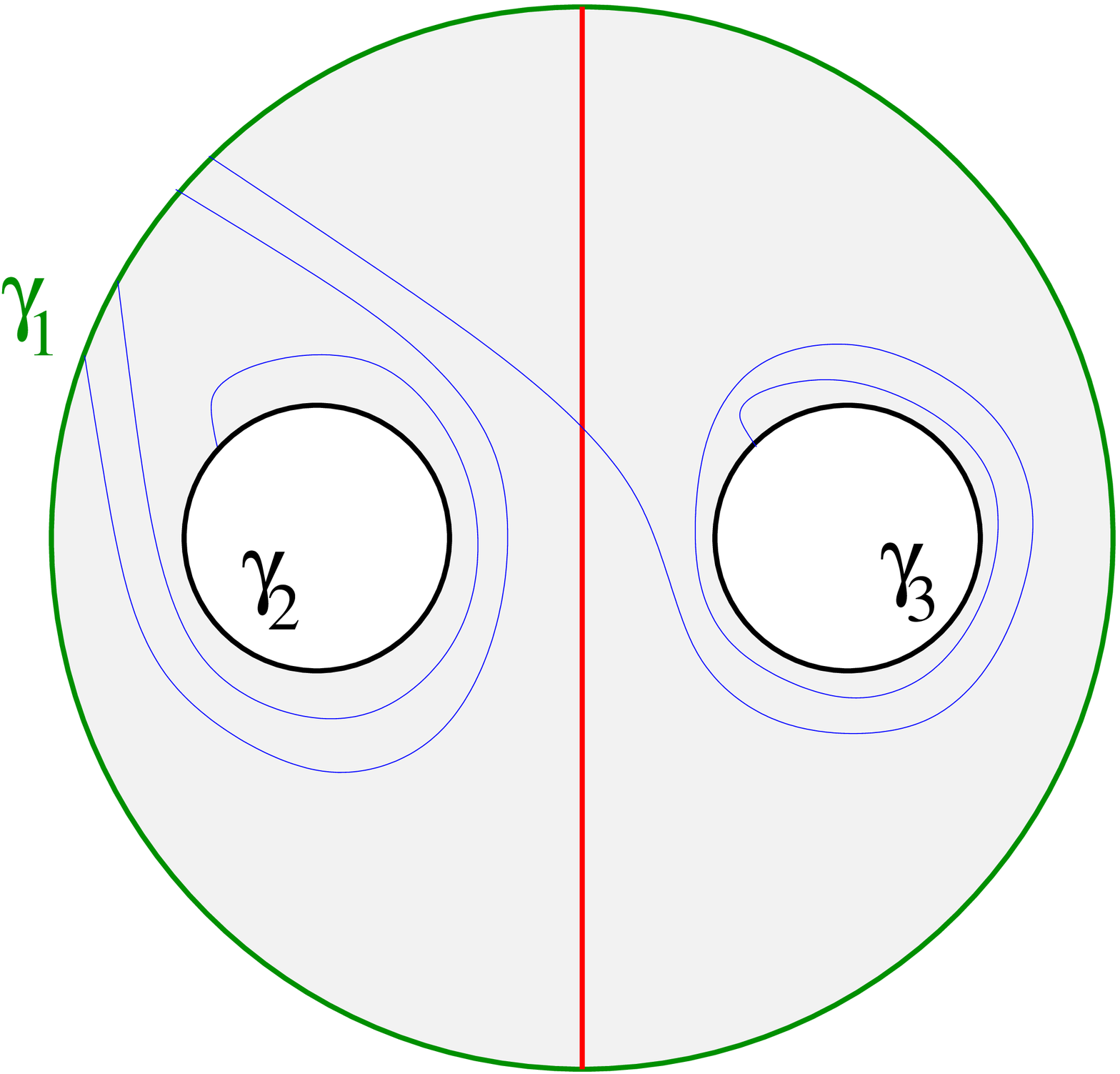}
\includegraphics[scale=.27]{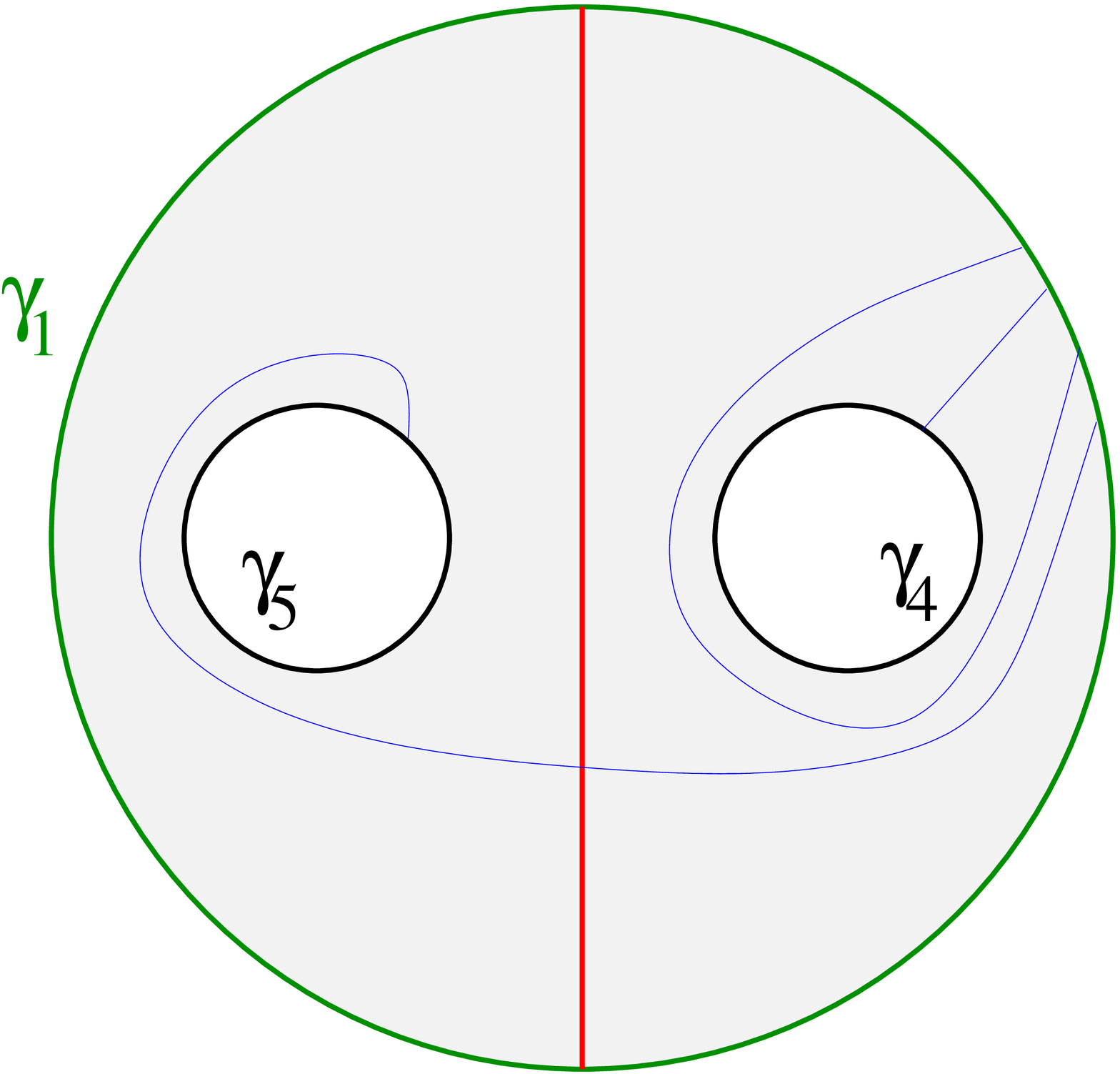}
\medskip

The top and bottom pairs of pants.

\medskip

\includegraphics[scale=.26]{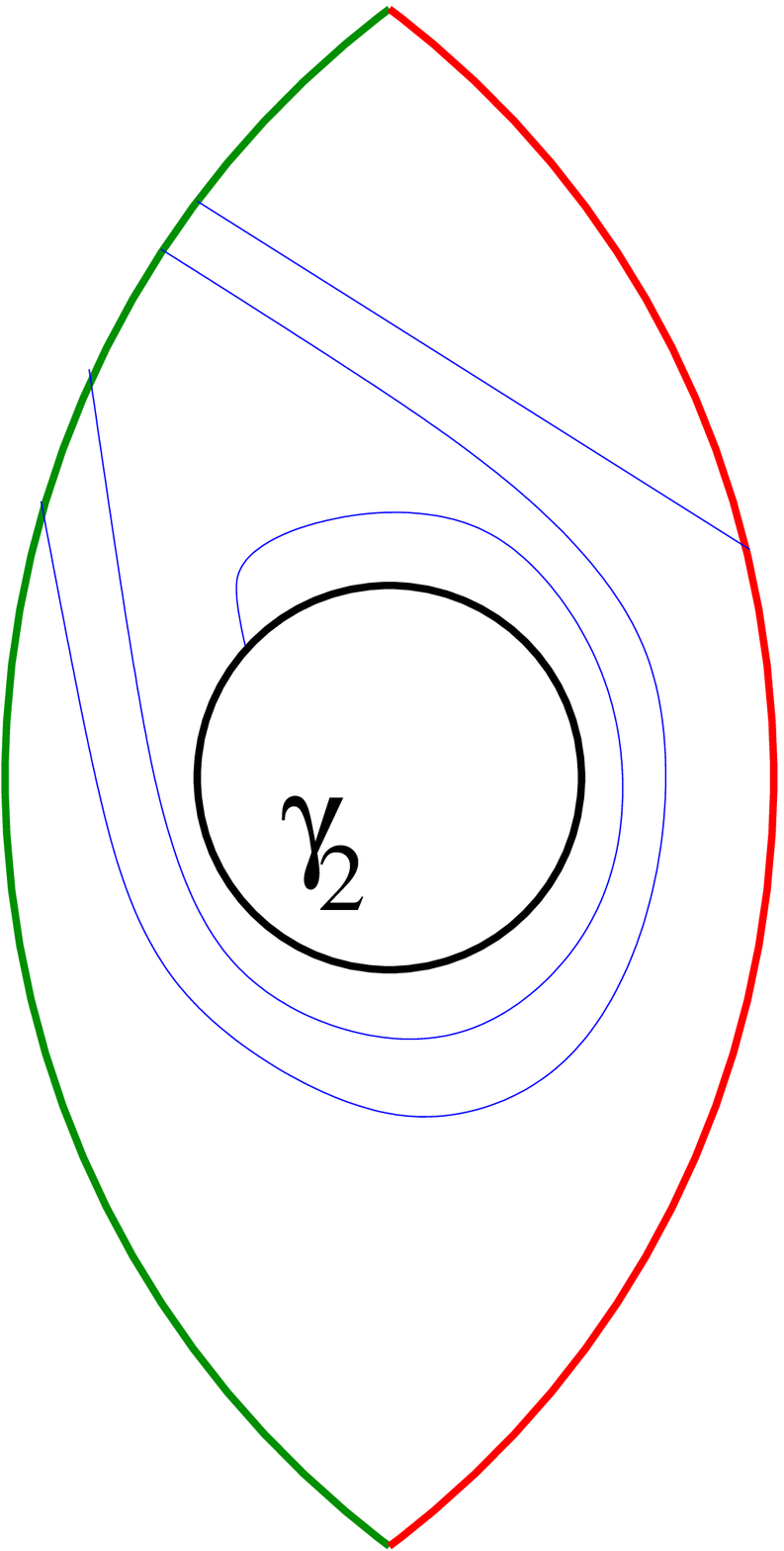}
\includegraphics[scale=.26]{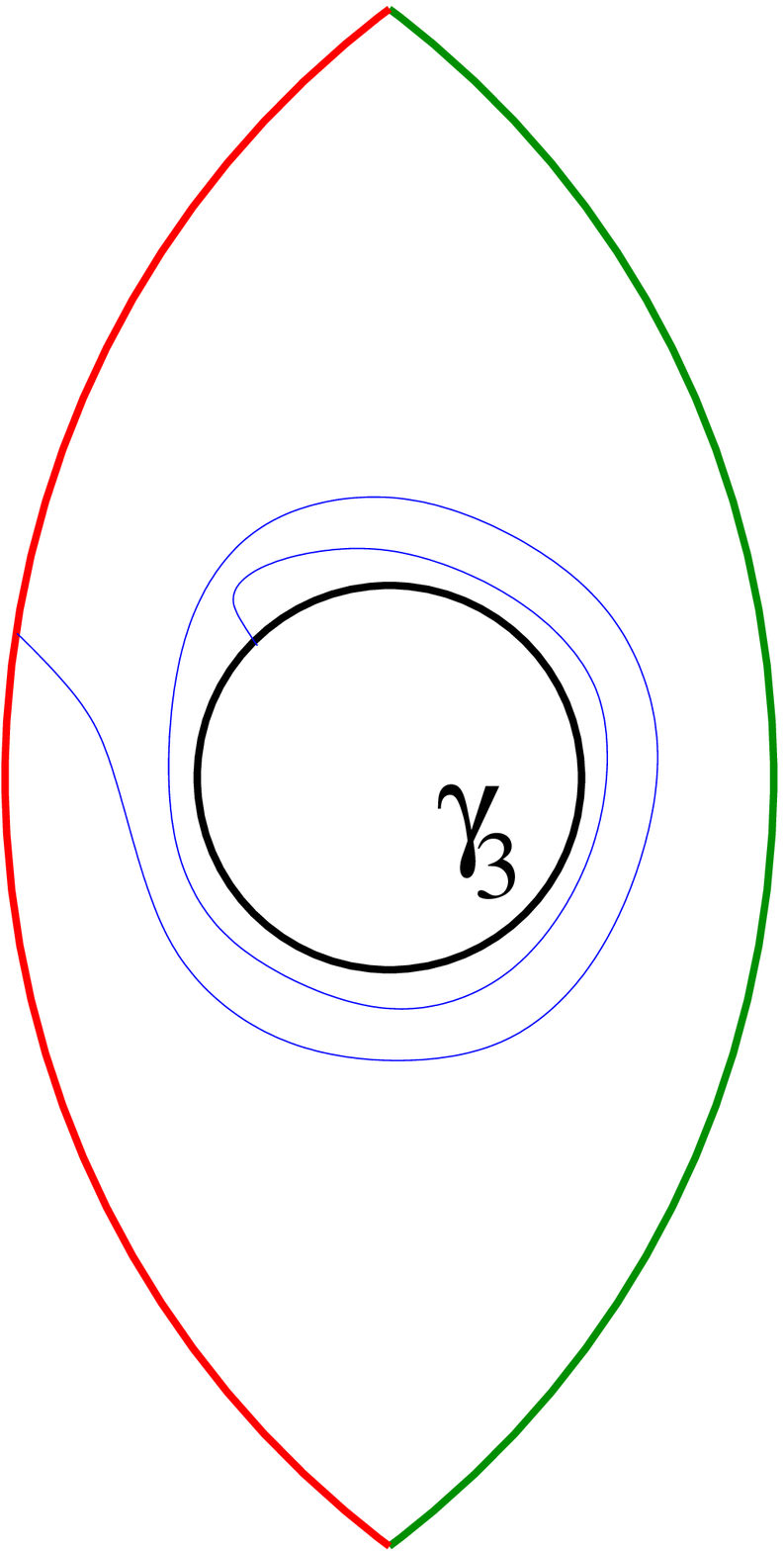}
\includegraphics[scale=.26]{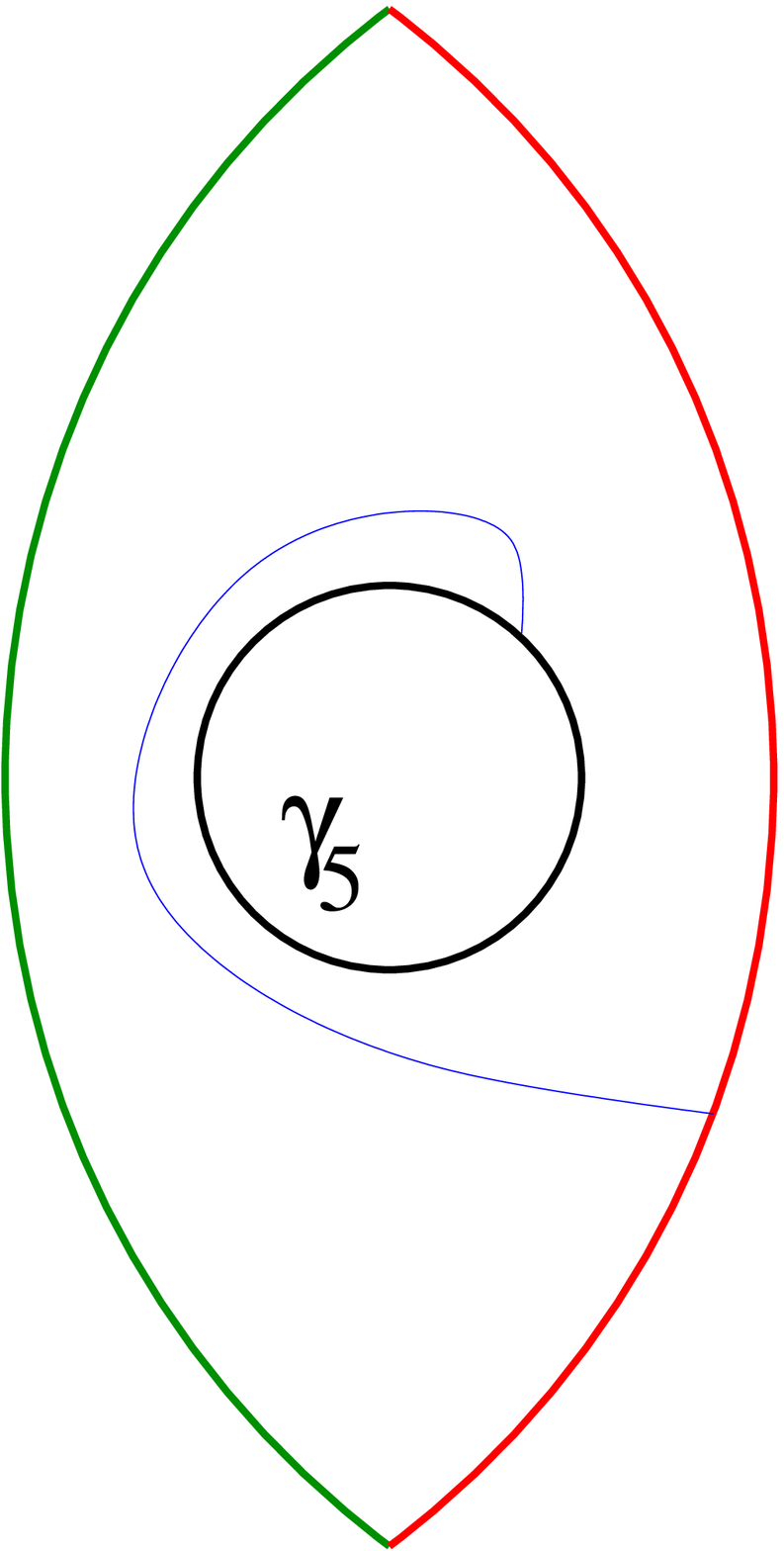}
\includegraphics[scale=.26]{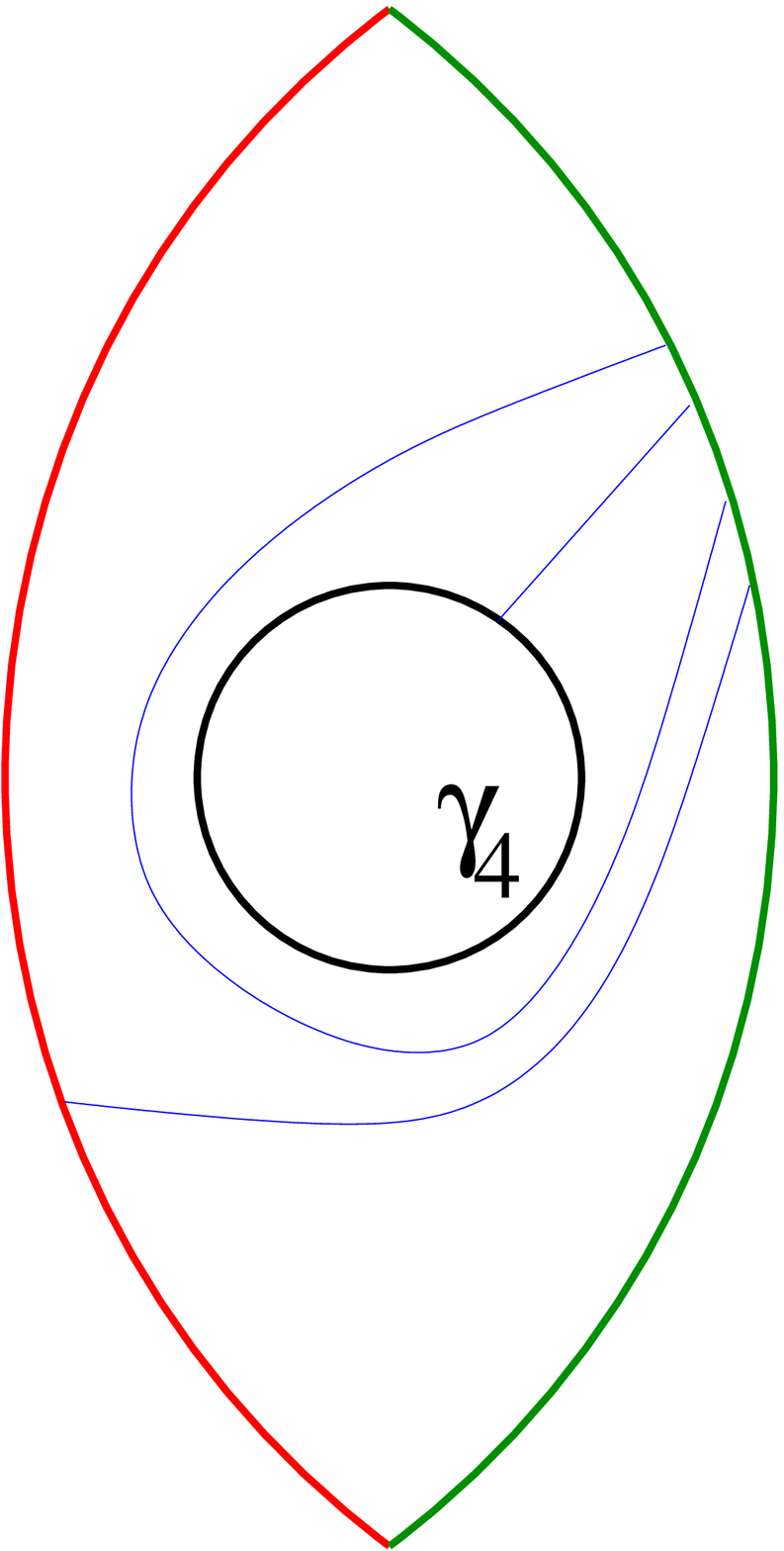}

\medskip

The four wedges.

\medskip

\includegraphics[scale=.27]{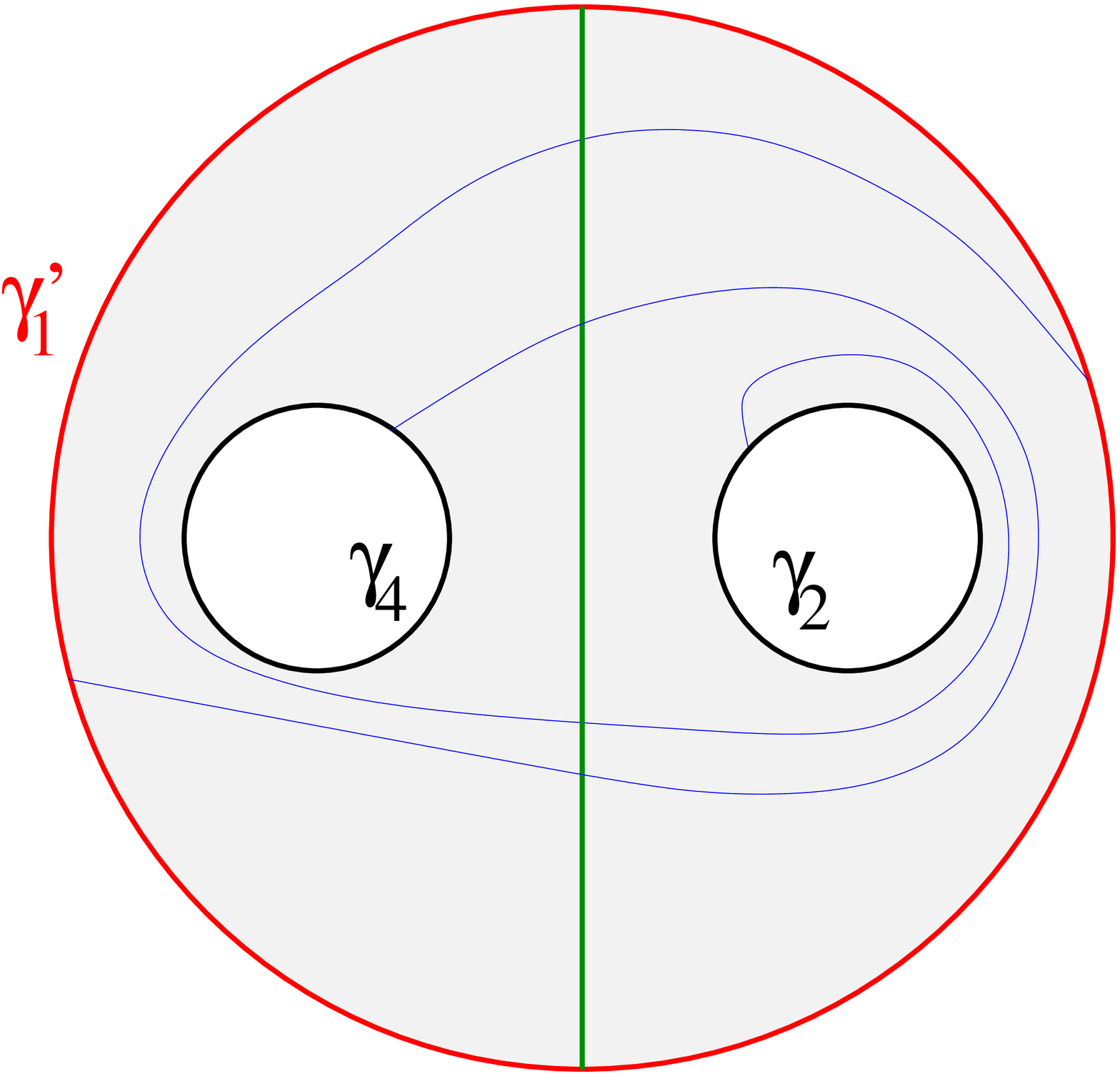}
\includegraphics[scale=.27]{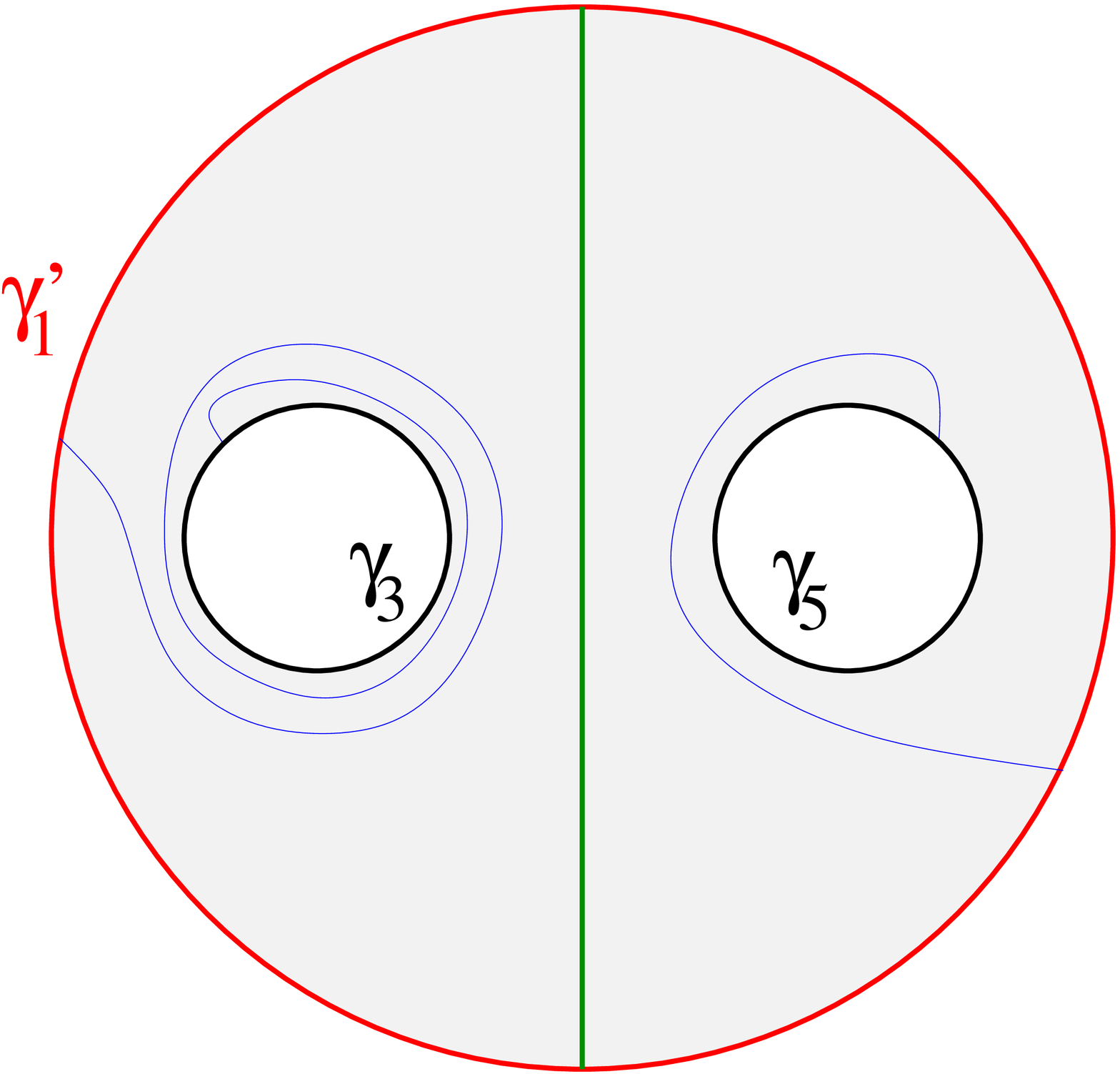}

\medskip

The left and right pairs of pants.

\caption{The transformation applied to $(4,1,1;-2,1,2)$.}
\label{big-picture}
\end{center}
\end{figure}

The action of the transformation on any particular geodesic can be
found as follows.  First cut the surface along $\gamma_1$, yielding a top 
pair of pants and a bottom pair of pants.  On each of these pairs of
pants, use Dehn's theorem to put the geodesic into standard form.
Next, cut each of the pairs of pants along a portion of
$\gamma_1'$, leaving four
wedges (one containing each of the other loops $\gamma_2$, \dots, $\gamma_5$).
These wedges can then be reassembled along the corresponding portions of
$\gamma_1$, giving
 a left pair of pants and a right
pair of pants, with common boundary curve $\gamma_1'$.  This is the 
decomposition ``after'' the Hatcher--Thurston transformation, and Dehn's
theorem can again be used to describe the parameters with respect to this
new decomposition.

We have illustrated this process in Figure~\ref{big-picture} for
a specific example.  We start with the geodesic whose Dehn--Thurston
parameters 
with respect to the lower quiver
in Figure~\ref{fig:SU(2)quivers}(b) are
$(p_1,p_2,p_3;q_1,q_2,q_3)=(4,1,1;-2,1,2)$.  In the top pair of pants, Dehn's
classification implies that we should have one arc of type $\ell_{12}$,
one of type $\ell_{13}$, and one of type $\ell_{11}$.  (Recall that this 
latter arc encirles $\gamma_2$.)  The twist parameters $q_2$ and $q_3$
are also applied in the top pair of pants, giving the curve depicted in
the upper left corner of Figure~\ref{big-picture}.

\begin{figure}[ht]
 \begin{center}
 \includegraphics[scale=.27]{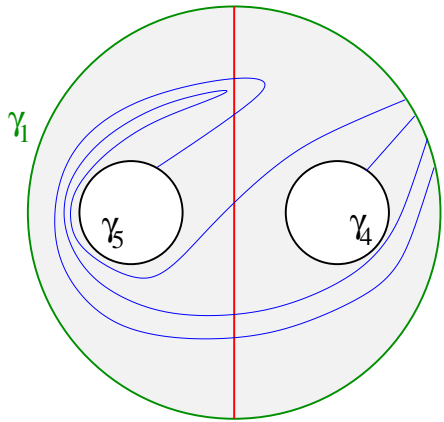}
 \caption{Another view of the bottom pair of pants.}
 \label{alt-bottom}
 \end{center}
 \end{figure}

In the bottom pair of pants, Dehn's classification is similar (since the
relevant $p_j$ parameters are identical): we should have one arc of type
$\ell_{14}$, one of type $\ell_{15}$, and one of type $\ell_{11}$ (which
in this case encircles $\gamma_5$).  We don't apply the twist parameters
$q_2$ and $q_3$ since they were already applied to the top pair of pants,
but we will apply $q_1$ here.  The clearest depiction of our curve is the
one given in Figure~\ref{alt-bottom}, in which all of this data can be
clearly seen.  However, for further manipulation, it is better to start
with a version in which the arcs have been ``pulled taut'', yielding the
curve depicted in the upper right corner of Figure~\ref{big-picture}.

Now the procedure is precisely as described above: we cut each of the 
pairs of pants from the first row of Figure~\ref{big-picture} in half,
giving the four wedges in the second row of Figure~\ref{big-picture}.
The attachment of the wedges is such that we can cyclically
permute them: the middle two of those wedges form the right pair of pants
shown in the lower right of Figure~\ref{big-picture}, while the first
and last wedge (joined in the opposite order, after the cyclic permutation)
form the left pair of pants shown in the lower left of 
Figure~\ref{big-picture}.

From the last row of Figure~\ref{big-picture} we can then read off the
Dehn--Thurston parameters with respect to the upper quiver in 
Figure~\ref{fig:SU(2)quivers}(b).  The left pair of pants contains one arc
of type $\ell_{1'2}$ and another of type $\ell_{1'4}$; it follows that
$p_1'=2$ and $p_2'=p_4'=1$.  Similarly, the right pair of pants contains
one arc of type $\ell_{1'3}$ and another of type $\ell_{1'5}$, so that
$p_1'=2$ and $p_3'=p_5'=1$.  The twist parameters can also be read off of
the bottom row: we have $q_1'=2$ (seen in the left pair of pants), $q_3'=2-1$
(seen in the right pair of pants) and $q_2'=0$.  (Notice that we had to
take the sum of the twisting around $\gamma_3$ and the twisting
around $\gamma_5$ in order to compute $q_3'$, since those curves are
attached on $\Sigma$.)  The full set of parameter
values is $(p_1',p_2',p_3';q_1',q_2',q_3')=(2,1,1;2,0,1)$.

This provides a new explicit prediction for the action of the S-duality 
group on this specific Wilson-'t~Hooft operator in the $\Ncal=2$ 
conformal theories 
with gauge group $SU(2)^3$ and no flavor symmetry. Other 
examples can be worked out in a similar way, and the general 
formulas can be found in the references \cite{MR1144770,MR743669}.

\section{Conclusions and discussion}
\label{sec-discussion}

We have classified all $1/2$ BPS loop operators in $\Ncal=2$
generalized quiver conformal field theories which are the IR limit
of two  coincident M5-branes wrapping an arbitrary Riemann
surface with a single type of puncture.
Each pants decomposition of the Riemann surface
corresponds to a 
particular duality frame, where we associate an $SU(2)$ gauge group
factor to each connected curve
along which the surface is cut, and an
$SU(2)$ flavor group to each puncture.

In space-time the loops were completely trivial, either a straight line
or a circle, and the classification involved only studying the gauge
degrees of freedom. 
One may place an arbitrary
't~Hooft loop in any gauge group and as discussed in
Section~\ref{sec-CFT} it is also natural to allow non-trivial bundles 
for
the flavor groups. In these theories there are hypermultiplets
transforming in the fundamental of three groups (gauge or flavor), and
in order to couple them consistently to the gauge bundles, one has to
impose the usual Dirac quantization condition.
The total magnetic charge felt by each such hypermultiplet should be
an even integer to ensure that the Dirac string is invisible. 
We can also include 
a Wilson loop in an arbitrary representation of each of the gauge
group factors.
More generally, by combining 't~Hooft and Wilson loops, we get
dyonic loop operators.

This construction gives two integers for each $SU(2)$ gauge symmetry
and one for each flavor $SU(2)$,
leading to a set of integers  (\ref{gen-weights2b}) subject to 
the Dirac condition (\ref{gen-even}).

Remarkably, this is exactly the same data as in the classification of
homotopy classes of non-self-intersecting curves on Riemann surfaces
(possibly disconnected).
For each way of cutting the surface into pairs of pants we have
an identification of each $SU(2)$ gauge symmetry
with a glued pair of pants-legs and an identification of the flavor
groups with external punctures. The number of lines crossing
a pants-leg is the magnetic charge in that group and the twist which
the lines perform there is the electric charge. Dehn's theorem
gives precisely the same set of data we have identified in the gauge 
theory. The exact map is given in Table~\ref{table-dic}.

\begin{table}[t]
\begin{center}
\begin{tabular}{|p{2.5in}||p{2.9in}|}
\hline
Gauge theory& Riemann surface\\
\hline 
& \\
[-1.2em]
\hline
$SU(2)$ gauge group & Regular pants-leg\\
\hline
$SU(2)$ flavor symmetry & Degenerate pants-leg (puncture)\\
\hline
$T_2$ theory (four  hypermultiplets)& Pair of pants\\
\hline
Gauge couplings and theta angles & Complex structure moduli\\
\hline
(Generalized)
Wilson-'t~Hooft loop with weights
$(p_j;q_j)$ 
&Non-self-intersecting curve with Dehn-Thurston
parameters $(p_j;q_j)$\\
\hline
't~Hooft loop in group $SU(2)_j$ with weight $p_j$& 
$p_j$ curves crossing $\gamma_j$\\
\hline
Wilson loop in group $SU(2)_j$ with charge $q_j$&
Twist $q_j$ on the lines crossing the $\gamma_j$, 
or a disconnected piece of the curve wrapping 
$q_j$ times  the curve $\gamma_j$\\
\hline
Non-dynamical monopole field of weight $p_j$ for flavor group $SU(2)_j$&
$p_j$ open lines ending at the $[j-(3g-3)]^{\text{th}}$ puncture\\
\hline
Weyl symmetry&The conditions: $p_j\geq0$ and if $p_j=0$ then $q_j\geq0$\\
\hline
S-duality frame & Choice of a pants decomposition\\
\hline
S-duality group & Mapping class group of the surface\\
\hline
\end{tabular}
\parbox{5in}{\caption{Dictionary relating loop operators and 
geodesics.
\label{table-dic}
}}
\end{center}
\end{table}

The S-duality group of the four-dimensional gauge theory is expected 
to be the mapping class group of the associated Riemann surface. 
Having a classification of loop operators in terms of the 
geometrical data allows us to make the following conjecture:
\begin{quote}
\textit{The action of the S-duality group on Wilson-'t~Hooft operators 
in the $\Ncal=2$ theories based on $SU(2)$ gauge factors 
is given by the action of the mapping class group on 
homotopy classes of
non-self-intersecting curves on the relevant Riemann surface.}
\end{quote}
Our classification of loop operators in these theories, 
combined with its identification with Dehn's classification of the 
curves and Penner's theorem on their transformation rules 
furnish (in principle) an explicit map for any such gauge theory.

We have illustrated this in several examples. In the case of a 
single gauge $SU(2)$ and $N_F=4$ as well as for the 
$\Ncal=2^*$ theory, this matches exactly the known expressions.
In other cases the general rules are quite complicated, but we 
demonstrated in several examples how to use 
Penner's algorithm to analyze the transformation rules and 
predict the action of the S-duality group on loop operators.

As pointed out in Section~\ref{sec-geodesics}, the classification 
of homotopy classes of non-self-intersecting curves 
is identical to
the classification of non-self-intersecting geodesics with
respect to any fixed hyperbolic metric. 
Geodesics with respect to hyperbolic metrics are
very natural 
and arise in the large $N$ holographic theory as discussed in 
Appendix~\ref{app-ads}.

Another way to construct these conformal field theories is in terms
of M5-branes  wrapped on a Riemann surface in space-time.
For any Riemann surface with an arbitrary metric
there exists a $U(1)$-invariant Ricci flat metric defined in a neighborhood
of the zero-section of the cotangent bundle on the surface
\cite{alg-geom/9710026,MR1817502}.
It should be possible to construct a supersymmetric
embedding of an M2-brane ending on the M5-branes. In the six
dimensional theory this corresponds to a Wilson surface, and
it would be interesting to follow the double dimensional
reduction to loop operators in four dimensions.
One ends up again with curves on the Riemann surface 
which now may have an arbitrary metric.%
\footnote{Note that if the metric 
on the Riemann surface does not have constant curvature, {\it i.e.,}
is not hyperbolic, 
the homotopy classes do not have as close a connection to
geodesics as they did for hyperbolic metrics. Still,
each homotopy class
of non-self-intersecting curves can be represented by a
shortest-length geodesic which is non-self-intersecting 
\cite{MR671777}, at least in the case without punctures.}
Perhaps it is possible to address 
from this point of view what happens when the curves
self-intersect, in particular why this is not allowed on
two M5-branes, but is acceptable in the large $N$ limit.

Yet another possible description of this system is in terms of 
brane webs.  One should
consider infinitely long $(p,q)$ strings and/or D3-branes 
ending on the 5-branes
whose world volume supports the $\Ncal=2$ theories \cite{Benini:2009gi}.

One obvious generalization is to consider Gaiotto's construction
of $\Ncal=2$ conformal field theories based on $SU(N)$ rather than
on $SU(2)$ (describing $N$ coincident
M5-branes) \cite{Gaiotto:2009we}.
There are two difficulties one encounters, the first
is just the richness of the construction, with many possible 
``quiver tails'' involving extra groups of rank less than $N$. 
The other difficulty with these theories is that the coupling between 
the different gauge factors is through
some strongly interacting theories, of which little is known.

Neglecting possible loop operators that might exist
intrinsically in the $T_N$ theory,
one can study the conditions that the Wilson-'t~Hooft operators constructed
from  gauge fields must satisfy.
An obvious requirement is the Dirac quantization condition.
An arbitrary 't~Hooft loop in each $SU(N)$ factor
is specified by a co-weight
$\mu$ which is the diagonal matrix
$\mu={\rm diag}(m_I-|m|/N)_{I=1}^N$, where $m_I$ are non-negative integers 
and $|m|=\sum_I m_I$.%
\footnote{Using a Weyl transformation we can assume the integers 
are ordered and then $m_I$ is the number of boxes in the $I^{\text{th}}$ row
of a Young tableau, which corresponds to a representation of the
Langlands dual of the Lie algebra 
$\mathfrak{su}(N)$. The Dirac condition then means that
the sum of $N$-alities of the three representations is trivial.}
In each $T_N$, there is an operator $Q_{a_1a_2a_3}$
that transforms in the tri-fundamental representation 
under the symmetry group $SU(N)_1\times SU(N)_2\times SU(N)_3$
of $T_N$.
For $Q_{a_1a_2a_3}$ to be single-valued around the Dirac string,
we require that the sum $|m^{(1)}|+|m^{(2)}|+|m^{(3)}|$ be divisible by $N$,
where $|m^{(j)}|$ is the quantity $|m|$ associated to $SU(N)_j$ for $j=1,2,3$.

Adding also arbitrary Wilson loops (consistent with the 't~Hooft loops)
leads to $2N-2$ integers for each gauge group factor subject to the Weyl
symmetry and the Dirac quantization condition. The flavor groups, 
``quiver tails'', and possibly the $T_N$ factors involve extra data. 
Unfortunately, for $N>2$ we do not 
currently have an analogous geometric classification of curves that 
matches this data. We hope to address this issue in future
work \cite{N>2}.

For very large $N$ we can go to strong
't~Hooft coupling and study the theory using its M-theory dual, where 
the loop operators are given by arbitrary geodesics on the Riemann 
surface with the hyperbolic metric, allowing for self-intersections 
(see Appendix~\ref{app-ads}).
The reason self-intersections are allowed is that the M2-brane does 
not really cross itself, but is separated along another $S^1$ in
the geometry. This is in contrast to the case of $SU(2)$ where
crossings were not allowed. For finite $N$ we might expect a 
``stringy exclusion principle'' on this $S^1$ to somehow restrict the 
possible crossings on the surface.

Thus far we have only discussed the classification of loop operators,
but a very natural question to ask is what their expectation value is.
For this purpose one should focus on the case of the loop with 
a circular geometry in space-time, 
which
still preserves global supersymmetry, but whose expectation value
is non-trivial.
Note also that the expectation value should depend on the gauge couplings,
and hence on the complex structure moduli of the Riemann surface.  

Using the supergravity dual discussed in Appendix~\ref{app-ads}, 
the result for large $N$ and strong
coupling is very simple to write down. 
The (Euclidean) M2-brane solution has 
the geometry $\HH_2\times S^1$ (assuming it is connected). 
The $S^1$ is related to a geodesic on 
the Riemann surface, which now has a hyperbolic metric 
and can be written as 
$\HH_2/\Gamma$ with $\Gamma$ a discrete subgroup of
$SL(2,\R)$ (and the two $\HH_2$ should not be confused). 
Any connected geodesic can be related to an element
$\gamma\in\Gamma$
and the length of the geodesic is the norm of $\gamma$ times 
the unit length on the Riemann surface.
Multiplying the M2-brane tension $T_{M2}=1/4\pi^2$ it
appears as the effecting string tension in $AdS_5$. For the
$\HH_2\subset AdS_5$ the standard calculation of the area gives
$-2\pi$
(after canceling a divergent contribution by a boundary term) 
\cite{Beren-Corr,DGO}.
In the metric (\ref{MN-metric}), the radius of $AdS_5$ 
is $2^{1/2} (\pi N)^{1/3}$,
and the radius of $\HH_2/\Gamma$ is $(\pi N)^{1/3}$.  So the vacuum 
expectation value is
\be
\left<L_\gamma\right>
\sim\exp |\gamma|N\,.
\ee
The gauge coupling appears in 
$|\gamma|$, since it depends on the moduli of the surface, which are 
identified with the gauge couplings. 
In particular, if the surface 
has a very short geodesic, this corresponds to small gauge coupling 
and we expect the length of the geodesic to be the gauge coupling 
$g_\text{YM}^2$, giving
\be
\left<L_\gamma\right>
\sim\exp g_\text{YM}^2N\,.
\ee
This dependence is different than for $\Ncal=4$ SYM, where 
for the circular Wilson loop
\be
\left<L_\gamma\right>_{\Ncal=4}
\sim\exp \sqrt{g_\text{YM}^2N}\,.
\ee

From the gauge theory side, an
exact calculation using localization was done by
Pestun in \cite{Pestun:2009nn} which expresses the expectation
value of a Wilson loop as a modification of the Gaussian matrix model 
of \cite{Erickson, Dru-Gross} 
by the Nekrasov partition function. This formula applies only for
Wilson loops, but if one assumes S-duality, then some 't~Hooft or
dyonic loops can be
transformed into purely electric ones. 
In fact, if we identify loop operators with
geodesics on a Riemann surface, then each non-self-intersecting
geodesic is part of a maximal set of disjoint non-self-intersecting
geodesics which can be used to cut the surface into pairs of pants and
can thus be declared the Wilson loops 
in an appropriate duality frame.
Therefore assuming S-duality, Pestun's formula calculates all circular loop
operators for generalized quivers based on $SU(2)$. 

In order to check our prediction of the S-duality transformation rules, 
it would be extremely interesting to have an independent 
calculation of the expectation 
values of 't~Hooft operators, or more general loop operators. 
One should be able to repeat the calculation made in the $\Ncal=4$ super 
Yang-Mills case \cite{Gomis:2009ir,Gomis:2009xg} or perhaps generalize the 
localization argument in \cite{Pestun:2009nn} to general loop 
operators.

In fact very
recently it was pointed out that in the absence of any loop operators, when Pestun's
formula gives the partition function on $S^4$, the expression
can be reinterpreted as a correlation function in Liouville theory 
\cite{Alday:2009aq}.
Using our classification of loop operators it should be possible to
extend the correspondence also to the full formula including the
expectation value of Wilson loops (and by S-duality to all loop operators).

\section*{Acknowledgments}

We are grateful to Fernando Alday, Francesco Benini, Sergio Benvenuti, 
Daryl Cooper, Harald Dorn, Mike Freedman,
Davide Gaiotto, Jaume Gomis, Sergei Gukov, Nick Halmagyi,
George Jorjadze, Anton Kapustin, Darren Long, Juan Maldacena, 
Ronen Plesser, and J\"org Teschner.
We would like to thank the Kavli Institute for Theoretical Physics,
where this project was initiated, for its hospitality.
N.D. and T.O. would like to thank the Benasque Science Center
for its hospitality.
Research at the Perimeter Institute is
supported in part by the Government of Canada through NSERC and by
the Province of Ontario through MRI.
This research was
supported in part by the National Science Foundation under Grant No.\ 
PHY05-51164 and DMS-0606578.
Any opinions, findings, and conclusions or recommendations expressed in this 
material are those of the authors
and do not necessarily reflect the views of the National Science Foundation.


\appendix

\section{M2-branes as holographic loop operators}
\label{app-ads}

To complement the gauge theory analysis in the body of this paper,
we consider the analogous problem for gauge theories based on the group
$SU(N)$ in the large $N$ limit and at strong coupling, for which there is
a dual supergravity description through the $AdS$/CFT correspondence. It is quite
straightforward and unambiguous to construct Wilson, 't~Hooft, and
dyonic loops in the supergravity theory which is instructive to 
compare with the gauge theory and topology discussed in the main text. This
appendix can be read independently from the rest of the paper, and likewise
it can be omitted by the reader interested solely in the gauge theory.

A related gauge theory is 
$\Ncal=4$ super Yang-Mills which is dual to type IIB on $AdS_5\times S^5$,
the simplest example of the $AdS$/CFT correspondence. 
In this gauge theory, 
the most symmetric Wilson and 't~Hooft operators follow a straight line or
a circle in $\R^4$. A Wilson loop is described at strong coupling by
a fundamental string and an 't~Hooft loop by a D1-brane; in both
cases they occupy an $AdS_2$ subspace of $AdS_5$.
While in the gauge theory Wilson loops and 't~Hooft loops are very
different objects (in particular, the latter has no simple description as an insertion
of operators made of the electric variables), in string theory the difference
is just in the choice of brane probe. The distinction between the two gets
even smaller when we go back to the M-theory picture, and use
a $T^2$ in place of the hyperbolic Riemann surface. Now both the
fundamental string and
D1-brane are M2-branes wrapping some cycle inside $T^2$ and
two directions ($AdS_2$ or $H_2$) in the non-compact space.
The distinction is simply which cycle
they wrap: the one associated with the ``11th direction''
gives rise to a fundamental string, another to a D1-brane,
and a generic one to a $(p,q)$ string.

The situation for more general Riemann surfaces is quite analogous. One
considers an M2-brane occupying an $AdS_2\subset AdS_5$ while
following a curve on the Riemann surface.%
\footnote{This fact was already mentioned in \cite{Gaiotto:2009gz}.}
As we show below, 
to preserve supersymmetry, this curve has to be a geodesic. 
The electric and
magnetic charges of the loop operator under the different gauge groups can be
read from the way the geodesic wraps or crosses the corresponding necks
on the surface, in exact analogy to the relation between curves and loops 
described in the main text.

This leads to the identification:
\begin{quote}
\textit{The classification of maximally supersymmetric Wilson-'t~Hooft loop
operators
in $\Ncal=2$ conformal generalized quiver theories at large $N$ is
given by arbitrary geodesics (possibly self-intersecting) 
on the relevant Riemann surface.}
\end{quote}
Note that the only distinction from the case of gauge group factors of 
$SU(2)$ studied in Sections~\ref{sec-CFT} and~\ref{sec-geodesics} 
is that for large $N$ self-intersections of the geodesics are allowed.

The gravity duals of the $\Ncal=2$ conformal theories are obtained
as the back-reaction of $N$ M5-branes wrapping a 
hyperbolic Riemann surface $\Sigma$ \cite{Gaiotto:2009gz}.
We restrict our analysis to the case of a Riemann surface without punctures, 
where the gravitational background is the Maldacena-Nu\~nez solution
\cite{Maldacena:2000mw} given by
\be
ds_{11}^2
=(\pi N l_p^3)^{2/3}\f{W^{1/3}}{2}
\bigg[
4 ds^2_{AdS_5}+2ds^2_{\Sigma}
+2d\theta^2+\f{2}{W}\cos^2\theta ds^2_{S^2}
+\f{4}{W} \sin^2\theta (d\chi+v)^2
\bigg].
\label{MN-metric}
\ee
Here $ds^2_{\Sigma}$ is the metric on $\Sigma$ with
constant scalar curvature $-2$, 
$ds^2_{S^2}$ is the unit metric on the two-sphere 
and $W=1+\cos^2\theta$.
The metric contains a particular one-form $v$
on $\Sigma$ which we will discuss below.

We now wish to find explicit embeddings of M2-branes
that preserve a maximal amount of supersymmetries and represent loop operators.
Such loop operators should preserve an $SO(2,1)\times SO(3)\times SO(3)$
subgroup of the isometry group.
We thus assume that the world-volume spans an $AdS_2$
subspace and sits at the end point $\theta=\pi/2$
of the interval $0\leq \theta\leq \pi/2$.
Then the M2-brane wraps a closed curve $\tilde C$
in the three-dimensional space in which the circle parametrized
by $\chi$ is fibered over $\Sigma$.
The projection of $\tilde C$ defines a closed curve $C$ on $\Sigma$.

To study supersymmetry we need eleven-dimensional
Killing spinors satisfying the equation
\be
\nabla_m \eta+\oo{288}\left[\Gamma_{m}^{~~npqr}-8\delta^{n}_m
\Gamma^{pqr}\right] G_{npqr}\eta=0.
\ee
The metric (\ref{MN-metric}) is a special
case of the more general metrics found in \cite{Lin:2004nb},
which characterize $\Ncal=2$ superconformal field theories in four dimensions.
We can thus adapt the Killing spinors obtained there for
the background (\ref{MN-metric}).%
\footnote{More precisely, the gravity solutions
with four-dimensional $\Ncal=2$ superconformal symmetries
were obtained by analytic continuation
from the solutions that describe
local operators in six dimensional $(2,0)$ theories.
The latter solutions were
found by explicitly constructing Killing spinors.
}
There are eight linearly independent Killing spinors
\be
\eta^{\alpha A}\,,\qquad
\eta^{c\dot\alpha}_{~A}
\qquad
(\alpha=1,2,\quad\dot\alpha =\dot 1,\dot 2,\qquad A=1,2)
\ee
corresponding to Poincar\'e supercharges
$Q^{\alpha A},~\bar Q^{\dot\alpha}_{~A}$
on the boundary.%
\footnote{Note that $\alpha, \dot\alpha$, and $A$
label eight spinors, each of which has 32 components.
Also $c$ indicates charge conjugation
and is not an index.}
Here $\alpha$ and $\dot\alpha$ are
left and right handed spinor indices, while $A$ is for the $SU(2)$ R-symmetry.
We  use the anti-symmetric tensors $\varepsilon^{AB},
~\varepsilon^{\alpha\beta},~
\varepsilon^{\dot\alpha\dot\beta}$ with $\varepsilon^{12}=1$
to raise and lower indices.

Let us decompose the eleven-dimensional gamma matrices
satisfying
\be
\{\Gamma^m,\Gamma^n\}=2\eta^{mn},~~m,n=0,1,\ldots,10
\ee
as%
\footnote{To keep equations simple,
we keep implicit the distinction between coordinate and frame indices.
}
\be
\begin{aligned}
AdS_5&:&&\Gamma^\mu=\gamma^\mu\otimes \gamma_{(2)}\otimes
\gamma_{(4)},&&\mu=0,1,2,3,4,
\\
S^2&:&&\Gamma^{5,6}=1_4\otimes \gamma^{5,6}\otimes \gamma_{(4)},
\\
\Sigma,~\chi,~\theta&:&&\Gamma^i=1_4\otimes 1_2\otimes\gamma^i,
&&
i=7,8,9,10.
\end{aligned}
\ee
$\gamma_{(2)}$ and $\gamma_{(4)}$ are chirality matrices 
with eigenvalues $\pm 1$. 
The Killing spinors can be expressed in terms of lower dimensional Killing
spinors as
\ba
\eta^{\alpha A}&=&
e^{\tilde\lambda/2}
\psi^\alpha\otimes\left[
(1+\gamma_{(2)}\otimes \gamma_{(4)})\cdot \chi^A\otimes e^{\f{\zeta}2\gamma_9}
e^{i\chi/2}\epsilon_0\right],
\\
\eta^{c\dot\alpha}_{~ A}&=&
e^{\tilde\lambda/2}
\psi^{c}_{\dot\alpha}\otimes\left[
(1-\gamma_{(2)}\otimes \gamma_{(4)})\cdot \chi_A\otimes e^{-\f{\zeta}2\gamma_9}
e^{-i\chi/2}\gamma_7\epsilon_0\right],
\ea
where $\zeta$ is determined by $y=- e^{3\tilde \lambda} \sinh\zeta$
in terms of the quantities $y$ and $\tilde\lambda$
defined in \cite{Gaiotto:2009gz}.
The fixed four-component spinor $\epsilon_0$
satisfies
\be
(i\gamma^9\gamma_{(4)}+1)\epsilon_0=(1-i\gamma^7\gamma^8)\epsilon_0=0.
\ee
We have introduced $AdS_5$ Killing spinors $\psi^\alpha$
satisfying the equation
\be
D_m \psi^\alpha=\half \gamma_m \psi^\alpha,~~~m=0,1,\ldots, 4,
\ee
and $\psi^c_{\dot\alpha}$  satisfying
\be
D_m \psi^c_{\dot\alpha}=-\half \gamma_m \psi^c_{\dot\alpha},
\qquad
m=0,1,\ldots, 4.
\ee
The two $S^2$ Killing spinors
$\chi^A$ satisfy
\be
D_m \chi^A=\f{i}2 \gamma_m \chi^A,
\qquad
m=5,6.
\ee

By writing the $AdS_5$ metric in terms of Poincar\'e coordinates as
\be
ds^2=\f{dx^\mu dx_\mu+dz^2}{z^2}
\ee
with $\mu=0,1,2,3$ and $z\equiv x^4$,
the $AdS_5$ Killing spinors can be explicitly given as
\be
\psi^\alpha=z^{-1/2}\psi^\alpha_{(0)},
\qquad
\psi^c_{\dot \alpha}
=z^{-1/2}
\psi_{(0)\dot\alpha},
\ee
where $\psi^\alpha_{(0)}$ and $\psi_{(0)\dot\alpha}$ are
constant spinors
\be
\psi^1_{(0)}=
\begin{pmatrix}
1\\
0\\
0\\
0
\end{pmatrix},
\qquad
\psi^2_{(0)}=
\begin{pmatrix}
0\\
1\\
0\\
0
\end{pmatrix},
\qquad
\psi^c_{(0)\dot 1}
=
\begin{pmatrix}
0\\
0\\
1\\
0
\end{pmatrix},
\qquad
\psi^c_{(0)\dot 2}
=
\begin{pmatrix}
0\\
0\\
0\\
1
\end{pmatrix}.
\ee
We represent the $AdS_5$ gamma matrices as
\be
\gamma^\mu=
\begin{pmatrix}
0&\sigma^\mu\\
\bar\sigma^\mu&0
\end{pmatrix},
\qquad
\gamma^4=
\begin{pmatrix}
1&0\\
0&-1
\end{pmatrix}.
\ee

The linear combination of supercharges
$Q=\xi_{\alpha A} Q^{\alpha A}
+\bar \xi_{\dot \alpha}^{~ A} \bar Q^{ \dot\alpha}_{~~ A}$
on the boundary corresponds to the spinor
\be
\eta=\xi_{\alpha A}\eta^{\alpha A}
+\bar \xi_{\dot \alpha}^{~ A} \eta^{c \dot\alpha}_{~~ A}.
\ee
in the bulk.
In our ansatz for the world-volume of an M2-brane, we have
\be
x^1=x^2=x^3=0\,,
\qquad
\theta=\pi/2\,.
\ee
The condition for supersymmetry  \cite{Becker:1995kb}
\be
\oo{3!} \varepsilon^{abc}
\Gamma_{mnp}\p_a X^{m}
\p_b X^{n}
\p_c X^{p}\eta=-\eta
\ee
reduces, in the static gauge, to
\be
\Gamma_{04}\Gamma_{\tilde C}\eta=-\eta\,,
\label{M2BPS1}
\ee
where $0$ and $4$ are the $AdS_2$ directions, and
\be
\Gamma_{\tilde C}=  t^i\Gamma_i
\ee
is the gamma matrix in the direction of the curve $\tilde C$.
Let us decompose the  tangent vector $t$ with respect to
the orthonormal frame:
\be
t=t^7 e_7+t^8 e_8+t^9 e_9.
\ee
By collecting terms with the same eigenvalues of $\gamma^4$ in (\ref{M2BPS1}),
one finds that
\be
-\bar \xi^{\dot\alpha}_{~ A}
\gamma_7 \epsilon_0
+(\bar\sigma_0 \xi_A)^{\dot\alpha} t^i \gamma_i
e^{i\chi} \epsilon_0=0\,,
\qquad
\xi_{\alpha A} \epsilon_0
+(\sigma_0 \bar \xi_A)_\alpha e^{-i\chi} t^i\gamma_i\gamma_7\epsilon_0=0,
\ee
or equivalently
\be
\begin{gathered}
\left[-\bar \xi^{\dot\alpha}_{~ A}
+(\bar\sigma_0 \xi_A)^{\dot\alpha} (t^7 -i t^8 +t^9 \gamma_9\gamma_7 )
e^{i\chi} \right]
\gamma_7
\epsilon_0=0\,,
\\
\left[\xi_{\alpha A}
+(\sigma_0 \bar \xi_A)_\alpha e^{-i\chi}
(t^7 +i t^8+ t^9\gamma_9\gamma_7)\right]\epsilon_0=0,.
\label{M2BPS2}
\end{gathered}
\ee
Since $\gamma_9\gamma_7$ changes the eigenvalue of $\gamma_7\gamma_8$
for which $\epsilon_0$  is an eigenvector,
we need that
\be
0=t^9\propto \dot \chi+ v_7 \dot x^7+v_8 \dot x^8, \label{t9}
\ee
where the dot indicates the derivative with respect to
the proper length.
Then (\ref{M2BPS2}) implies that
\ba
&&\bar\xi^{\dot\alpha}_{~A}=e^{i\chi_0}
(\bar\sigma_0 \xi_A)^{\dot\alpha},
\label{M2susy}
\\
&&
{\rm arg}(t^7+it^8)=\chi-\chi_0
\label{M2BPS3}
\ea
with $\chi_0$ being a constant.
It is natural to introduce a complex coordinate $w=r e^{i\beta}$
on $\Sigma$, in terms of which the metric and the one-form $v$ are given by
\be
ds^2_\Sigma=4\f{dr^2+r^2d\beta^2}{(1-r^2)^2}=\f{4 d w d\bar w}{(1-|w|^2)^2}
\,,
\qquad
v=\f{2r^2d\beta}{1-r^2}\,.
\label{Sigma-metric}
\ee
Then any real vector field $V=V^i \p_i$ ($i=7,8$)
with unit norm takes the form
\be
V=\f{1-|w|^2}2 e^{i \varphi} \p_w+c.c.,
\ee
and its covariant derivative is given by
\be
D V= D_i V^j dx^j\otimes \p_j
=i e^{i\varphi} \f{1-|w|^2}2
\left(d\varphi +\f{2r^2}{1-r^2}d\beta\right)\otimes \p_w
+c.c.
\label{DV}
\ee
If we take the tangent vector
$\dot x^i\p_i=\dot w \p_w +\dot {\bar w}\p_{\bar w} $
of the curve $C$ on $\Sigma$
to be of unit norm, (\ref{M2BPS3}) implies that ${\rm arg}(\dot w)
=\chi-\chi_0$.
Then from (\ref{t9}) together with (\ref{DV}) we find that
$\dot x^i \p_i$ is covariantly constant, {\it i.e.,}
\be
\ddot{x}^i+\Gamma_{ij}^k \dot x^i \dot x^j=0
\qquad
(i,j,k=7,8)\,.
\ee
Thus the curve $C$ has to be a geodesic.
The condition (\ref{t9}) implies that $\tilde C$ is also 
a geodesic in the total space of the circle fibration over $\Sigma$.

Conversely, one can lift any closed geodesic $C$ on $\Sigma$,
to a closed geodesic $\tilde C$ on the total space
of the circle fibration by setting
\be
\chi={\rm arg}(\dot w)+\chi_0
\label{last}
\ee
for some constant $\chi_0$.
The relation (\ref{last}) between the orientation of the 
tangent vector of $C$ and the position $\chi$ on the circle implies that 
even if the geodesic $C$ self-intersects, the 
M2-brane world-volume does not. 
This relation amounts to the vanishing of the last term in (\ref{MN-metric}) 
($v$ is exactly canceled by $d\chi$) which guarantees 
that the lengths of $C$ and $\tilde C$ are equal.

We conclude that for an arbitrary geodesic $C$, an M2-brane wrapping 
the uplift $\tilde C$ and the $AdS_2$ subspace of $AdS_5$ 
is half BPS. The supercharges preserved by the M2-brane are determined by
(\ref{M2susy}).

\bibliography{n2loop}

\end{document}